%% file: sample-sigconf.tex
\pdfoutput=1
\documentclass[sigconf]{acmart}
\makeatletter
\renewcommand\@formatdoi[1]{\ignorespaces}
\makeatother

\usepackage{etex}
\renewcommand\footnotetextcopyrightpermission[1]{} % removes footnote with conference information in first column
\setcopyright{rightsretained}
\setcopyright{none}
\renewcommand\footnotetextcopyrightpermission[1]{}
\usepackage{footnote}
\makesavenoteenv{tabular}
\makesavenoteenv{table}

\usepackage{listings}
\AtBeginDocument{%
  \providecommand\BibTeX{{%
\normalfont B\kern-0.5em{\scshape i\kern-0.25em b}\kern-0.8em\TeX}}}

\usepackage{booktabs}

\usepackage{caption}
\usepackage{subcaption}
\usepackage{tikz,pgfplots}
\usepackage{listings}
\include{AlgorithmStyle}
\include{mytikz}

\usepackage{array}
\usepackage{makecell}
\newcolumntype{L}[1]{>{\raggedright\let\newline\\\arraybackslash\hspace{0pt}}m{#1}}
\newcolumntype{P}[1]{>{\arraybackslash}L{#1\columnwidth}}

\theoremstyle{plain}

\newtheorem{mydef}{Definition} 

\shortauthors{}

\begin{document}

%%
%% The "title" command has an optional parameter,
%% allowing the author to define a "short title" to be used in page headers.
\title{A characterisation of system-wide propagation in the malware landscape}

%%
%% The "author" command and its associated commands are used to define
%% the authors and their affiliations.
%% Of note is the shared affiliation of the first two authors, and the
%% "authornote" and "authornotemark" commands
%% used to denote shared contribution to the research.
\author{David Korczynski}
\email{david.korczynski@cs.ox.ac.uk}
\affiliation{%
  \institution{Department of Computer Science\\University of Oxford}
}

%%
%% By default, the full list of authors will be used in the page
%% headers. Often, this list is too long, and will overlap
%% other information printed in the page headers. This command allows
%% the author to define a more concise list
%% of authors' names for this purpose.

%%
%% The abstract is a short summary of the work to be presented in the
%% article.
\begin{abstract}
System-wide propagation is frequently observed in malware, and there are several resources, like blog posts and similar, that detail some of the techniques used. However, there is currently no thorough study on the subject at large, and the full extent of system-wide malware propagation remains unknown. In this paper, we perform a systematic study on many real-world samples to comprehensively characterise system-wide propagation within the malware landscape and the goal is to use detailed and precise analyses to derive high-level views. We achieve this by collecting a diverse set of malware samples, analyse them in our Minerva malware analysis framework and then extract vast amounts of statistics about the results. We use these results to provide an in-depth discussion centred on four main research questions. 
\end{abstract}

%%
%% The code below is generated by the tool at http://dl.acm.org/ccs.cfm.
%% Please copy and paste the code instead of the example below.
%%

%%
%% Keywords. The author(s) should pick words that accurately describe
%% the work being presented. Separate the keywords with commas.

%%
%% This command processes the author and affiliation and title
%% information and builds the first part of the formatted document.
\maketitle
\thispagestyle{empty}

\section{Introduction}
The number of malware samples we discover continues to grow year by year and the ability to automatically analyse these threats is a critical part of our defense. The underpinnings of malware analysis can broadly be categorised in the two broad categories of static and dynamic analysis. Static analysis analyses the malware without executing it and dynamic analysis relies on executing the given sample.  Dynamic analysis is typically performed by way of a malware sandbox, i.e. by executing the sample in an isolated and dedicated environment usually based on some form of virtualisation. The popularity of malware sandboxes have grown significantly in the last decade \cite{argos:eurosys06, Bayer2006, Yin:2007:PCS:1315245.1315261, Egele:2008:SAD:2089125.2089126, 10.1007/978-3-642-37300-8_9, Henderson:2017:DPW:3057931.3057958, DKOR} and they are attractive because they allow precise analysis of concrete malware executions, which can be used to circumvent various obfuscation and packing techniques that are typically deployed by the malware.  

However, malware often comes with several capabilities in order to avoid dynamic analysis and in recent years the concept of evasive malware has been a popular technique for malware authors.  One particular technique adopted by malware is the concept of system-wide malware execution. This refers to malware executing throughout the entire system by way of multi-process propagation, code injection, code-reuse attacks and alike. In essence, this form of execution paradigm complicates the task of precisely capturing the malware execution since the malware may execute in stealthy ways, e.g. within the execution-context of otherwise benign processes. The academic community has responded and made several efforts into studying the problem \cite{DBLP:conf/malware/BaraboschG14, 198415, DBLP:conf/dimva/BaraboschEG14, DKOR, DBLP:conf/dimva/BaraboschBDP17, YuheiKawakoya2019}. 

Although system-wide malware propagation has received attention from the academic community, there are currently no comprehensive studies on the subject and the scope of the problem at large remains unknown. Most resources are anecdotal examples of code injection techniques \cite{HASHEREZADECodeInjection, Injectopi} and blog posts by anti-malware researchers \cite{DridexV4Atombinging, endgame_blog}, and these mainly present results based on manual analysis of specific samples. This is an important missing gap of knowledge since systematic studies of large malware data sets are essential for anti-malware research. We rely on these studies to systematise our knowledge, design new experiments that verify our techniques and to explore new research avenues. 

In this paper, we aim to fill this gap by performing the first large-scale systematic study of system-wide malware propagation. To this end, we build a comprehensive statistics component into the Minerva malware analysis sandbox presented \cite{2019arXiv190809204K} and then use our framework to analyse and characterise a large and diverse set of malware samples. Our study is focused on three aspects of system-wide propagation. First, to understand the prevalence and diversity of system-wide malware propagation. Second, to understand the relationship between system-wide malware propagation and malicious activities. Third, to understand system-wide malware propagation evolution over time and inter-family characteristics of malware propagation strategy.

To reason about system-wide malware executions at large, we define the concept of a system-wide control-flow graph, which is a data structure that describes the entire system-wide malware execution. Effectively, this data structure combines the ability to identify multi-process execution \cite{DKOR} and capture execution waves \cite{2019arXiv190809204K} to give a complete overview of a given malware propagation.

We perform our study with a comprehensive set of malware samples that reflects a diverse and broad view of malware in the wild. The data set we collect has samples from many different families, many different malware types and samples that were discovered over a seven-year time-frame. In addition to this, our data set is completely balanced in terms of samples per family so as to avoid biases in the results.

In summary, this paper makes the following contributions:
\begin{itemize}
    \item We propose the concept of a system-wide control flow graph which captures intrinsic aspects of malware propagation in a given sample.
    \item We collect a data set with a broad range of malware samples that were discovered across the last seven years.
    \item We implement a comprehensive and rigorous statistics component that interprets the results of Minerva.
    \item We characterise system-wide malware propagation through a detailed and thorough study of the malware samples by analysing them in Minerva and gathering many different and insightful statistics.  
\end{itemize}

\section{Minerva background}
The work we present in this paper uses our malware analysis system Minerva described in earlier work \cite{2019arXiv190809204K}. In order to make this paper self-contained we will briefly cover in this section the core techniques that Minerva uses to analyse malware which are relevant for this paper. 

Minerva traces the malware execution based on dynamic taint analysis so as to capture the malware execution throughout the entire system and the specific purpose is to monitor malware execution in the context of multi-process propagation. In short, Minerva taints the malware when it is loaded into memory, including both code and data sections of the malware module, and then considers malware execution to be each instruction executed on the system that is made up of tainted memory. Furthermore, the output of each tainted instruction is also tainted, so as to capture dynamically generated code by the malware. In this way, Minerva captures malware execution without any hooks on common APIs for process propagation, e.g. \texttt{CreateRemoteThread}. This specific approachin Minerva was adopted from Tartarus \cite{DKOR}.  

In addition to capturing system-wide malware execution, Minerva also captures dynamically generated code by the malware using an information-flow model. This is used for automatic unpacking and capturing of shellcode, and the key idea behind the information-flow approach is to capture dynamically generated code independently of who wrote the code. Specifically, it is often the case that malware will force benign code to create dynamically generated code on its behalf, and the information-flow model will capture this, contrary to the models of dynamically generated code used in many traditional unpackers that only capture dynamically generated code explicitly by the malware \cite{Kang:2007:RHC:1314389.1314399, Dinaburg:2008:EMA:1455770.1455779, Bonfante:2015:CMS:2810103.2813627, DBLP:conf/malware/Korczynski16}. In this paper, we call each wave of dynamically generated code an ``execution wave''. 

Minerva is built on top of the QEMU \cite{Bellard:2005:QFP:1247360.1247401} full-system emulator and, more specifically, Minerva relies on the PANDA \cite{Dolan-Gavitt:2015:RRE:2843859.2843867} framework. In this paper we have made minimal additions to Minerva, however, we have built a large statistics-framework around the system that allows us to use the precise malware analysis offered by Minerva to extract large amounts of data. 

\section{System-wide propagation graph}
\label{sec:SystemWideControlFlowGrap}
In this section, we introduce the concept of a system-wide propagation graph (SPG). Intuitively, SPG combines the output of Minerva into one single data structure. This data structure, thus, captures intrinsic aspects of host-based malware propagation in terms of multi-process execution and dynamically generated code. Informally, the SPG is a directed graph where the nodes constitute execution waves, and the edges constitute transitions from one execution wave to another. Each node then contains meta information to capture useful properties, such as their particular process. Formally, the SPG is described as follows. 

\begin{mydef}
A system-wide propagation graph is a 3-tuple, SPG=(V, E, $\hat v$), where G=(V, E) is a directed weakly-connected graph such that each node is an execution wave, each edge is a control-flow transition, and $\hat v \in V$ is the entry point.
\end{mydef}

To capture the $SPG$ in practice, we must be able to identify all of the execution waves and the corresponding control-flow transitions for a given malware sample. Minerva already supports capturing execution waves and identifying control-flow transitions between execution waves across processes (adopted from Tartarus \cite{DKOR}), however, to capture control-flow transitions from waves internally within a process, we make an edge from the last instruction executed before the entry of a new execution wave. 

To reason quantitatively about malware propagation we build several definitions on the SPG and the general idea is to describe malware propagation techniques in more ways than only counting the number of processes and the number of waves. We give insights about the malware propagation topology by creating two definitions that capture the maximum depth of processes and waves and one definition that describes the width of the $SPG$. 

\begin{mydef} 
Given an SPG, G=(V, E, $\hat v$), the process-depth of PD(G) is the maximum number of different processes visited amongst all non-cyclic path through G, starting from $\hat v$.
\label{def:processdepth}
\end{mydef}

\begin{mydef}
Given an SPG, G=(V, E, $\hat v$), the wave-depth WD(G) is the size of the longest non-cyclic path through G, starting from $\hat v$.
\label{def:wavedepth}
\end{mydef}

\begin{mydef}
Given an SPG, G=(V, E, $\hat v$), the SPG-width $|G|$ is the maximum number of non-cyclic paths from the entry-point $\hat v$ to all of the leaf nodes in the graph.
\label{def:SCFGWidth}
\end{mydef}

Notice that Definition \ref{def:processdepth} and Definition \ref{def:wavedepth} are different from calculating the total number of processes and waves, respectively. The notion of depth in the malware execution is relevant because it encapsulates how deep into the malware execution a manual malware analyst must go before reaching the end of an injection/unpacking sequence, for example, in a debugger. The notion of SPG-width given in Definition \ref{def:SCFGWidth} aims to capture how broadly malware executes on a system, and, therefore, describes the number of execution paths in the malware propagation rather than how deep they are.  

\subsection{Example of system-wide propagation graph}
\begin{figure}
\begin{minipage}{0.5\textwidth}
\begin{adjustbox}{scale=0.55}
\centering
\begin{tikzpicture}
	%\node (wa) [wa] {wave1};

	\node (proc0wave0) [wave2] {\texttt{Malware.exe}\\\textbf{$P_0W_0$}};
	\path (proc0wave0.east)+(2.0, +0.0) node (p0w1) [wave2] {\texttt{Malware.exe}\\\textbf{$P_0W_1$}};
	
	\fill (proc0wave0.west)+(-0.7,0) circle (0.1cm);	
	\path (proc0wave0.west)+(-0.7,0) node (EOP) {};
	\path [draw, ->] (EOP) -- node [above] {} (proc0wave0);

	\path [draw, ->] (proc0wave0.east) -- node [above] {} (p0w1.180);

	\begin{pgfonlayer}{background}
		\path (proc0wave0.west |- p0w1.north)+(-1.0,0.5) node (a) {};
        \path (proc0wave0.south -| p0w1.east)+(+0.5,-1.5) node (b) {};
        \path (proc0wave0.east |- p0w1.east)+(+4.0,-1.0) node (c) {};	
	
		\path[fill=gray!05, rounded corners, draw=black, dashed] (a) rectangle (c);
	\end{pgfonlayer}

	\path (p0w1)+(0.0, -2.5) node (p1w0) [wave2] {\texttt{winver.exe}\\\textbf{$P_1W_0$}};
	\path (p1w0)+(-0.0, -2.0) node (p1w1) [wave2] {\texttt{winver.exe}\\\textbf{$P_1W_1$}};
	
	\path [draw, ->] (p0w1.south) -- node [below] {} (p1w0.north);
	\path [draw, ->] (p1w0.south) -- node [below] {} (p1w1.north);

    \path (p1w1)+(-0.0, -2.0) node (p2w0) [wave2] {\texttt{explorer.exe}\\\textbf{$P_2W_0$}};
    
	\path (p2w0)+(-5.0, +0.0) node (p3w0) [wave2] {\texttt{dwm.exe}\\\textbf{$P_3W_0$}};
	\path (p2w0)+(-5.0, -2.5) node (p4w0) [wave2] {\texttt{taskhost.exe}\\\textbf{$P_4W_0$}};
	\path (p2w0)+(+0.0, -2.5) node (p5w0) [wave2] {\texttt{cmd.exe}\\\textbf{$P_5W_0$}};
	\path (p2w0)+(4.5, -2.5) node (p6w0) [wave2] {\texttt{conhost.exe}\\\textbf{$P_6W_0$}};
    \path (p2w0)+(4.5, 0.0) node (p7w0) [wave2] {\texttt{dllhost.exe}\\\textbf{$P_7W_0$}};
	
%	\path (p1w0.west) +(-2.5,+1.3) node (title) {\textbf{explorer.exe}};	
	
	\begin{pgfonlayer}{background}
		\path (p1w1.west |- p1w0.north)+(-0.3, 0.2) node (g) {};
		\path (p1w1.east |- p1w0.east)+(+0.3, -2.8) node (h) {};
		
		\path[fill=gray!05, rounded corners, draw=black, dashed]  (g) rectangle (h);
	\end{pgfonlayer}

	%\path [draw, ->] (proc0wave1.220) -- node [below] {\textbf{SendNotifyMessage}} (injectioncatalyst.130);
	\path [draw, ->] (p1w1.south) -- node [below] {} (p2w0.north);
    \path [draw, ->] (p2w0.west) -- node [below] {} (p3w0.east); 
    \path [draw, ->] (p2w0.205) -- node [below] {} (p4w0.east);
    \path [draw, ->] (p2w0.south) -- node [below] {} (p5w0.north);
    \path [draw, ->] (p2w0.335) -- node [below] {} (p6w0.west);
    \path [draw, ->] (p2w0.east) -- node [below] {} (p7w0.west); 
	%\path [draw, ->] (proc0wave1.east) edge [bend left=90, dashed] node [above] {\textbf{SetWindowLong}} (injectioncatalyst.east);

	%\path [draw, ->] (injectioncatalyst.east) edge [bend left=60, dashed] node [above] {} (proc1wave0.east);

\end{tikzpicture}
\end{adjustbox}
\end{minipage}

\begin{minipage}{0.4\textwidth}
\begin{align*}
    G &= (V, E, \hat v) \\    
    V &= \{P_0W_0, P_0W_1, P_1W_0, P_1W_1, P_1W_2\} \\
    E &= \{\{P_0W_0, P_0W_1\}, \{P_0W_1, P_1W_0\}, \\
      & \qquad \{P_1W_0, P_1W_1\}, \{P_1W_1, P_2W_0\},\\
      & \qquad \{P_2W_0, P_3W_0\}, \{P_2W_0, P_4W_0\},\\
      & \qquad \{P_2W_0, P_5W_0\}, \{P_2W_0, P_6W_0\},\\
      & \qquad \{P_2W_0, P_7W_0\}\} \\
    \hat v &= P_0W_0\\
      PD(G) &= 4 \\
      WD(G) &= 6 \\
      |G| &= 5 \\
\end{align*}
\end{minipage}
\captionsetup{justification=centering,margin=1cm}
\caption{The system-wide propagation graph of Tinba malware sample.}
\label{fig:Chapter5TinbaSPG}
\end{figure}
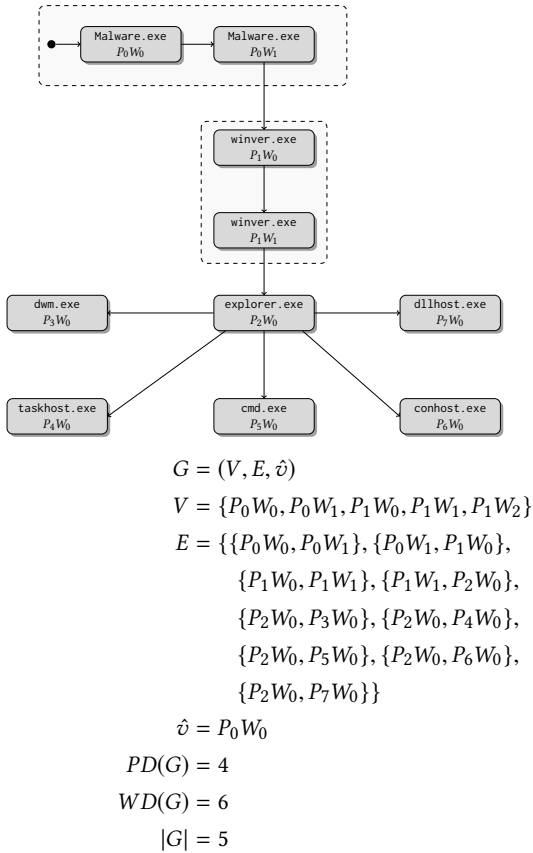

To illustrate system-wide propagation graphs, as well as process-depth, wave-depth and SPG-width, we show the SPG of a malware sample from the Tinba family in Figure \ref{fig:Chapter5TinbaSPG}. The sample initially creates one wave of dynamically generated code in its initial process and then launches a new process of the otherwise benign Windows application \texttt{winver.exe}. The malware injects code into the launched \texttt{winver.exe} process and the code in \texttt{winver.exe} further creates another wave of dynamically generated code. The second wave in \texttt{winver.exe} injects code into the otherwise benign Windows process \texttt{explorer.exe} and inside this process, the malware proceeds to inject code into five other processes on the system. The process-depth of the SPG is four, the wave-depth is six, and the SPG-width is five. 

\section{Research methodology}
\label{sec:chapter6ResearchMethodology}
We now elaborate on the methodology for our large-scale empirical study. Specifically, we outline our data set, our research questions and also describe our experimental setup, including false-positive removal.

\subsection{Data collection}
\label{sec:ChapterEmpiricalStudy_data_collection}
The goal of our data collection is to establish a set of samples that broadly represents the Windows malware landscape over several years. To this end, we have collected samples from various kinds of malware, a diverse set of families and from an extended timeline. We identify the year a sample is from as the year the sample was first submitted to VirusTotal. In contrast to other quantitative large-scale studies \cite{DBLP:conf/leet/BayerHBK09, Ugarte-pedrero_sok:deep} that analyse tens and hundreds of thousands of malware samples, we instead keep the total samples in our data set comparatively low and comprehensible but with a wide distribution of malware families. This is because samples within the same family tend to behave similarly \cite{CybIN}, and by keeping the numbers low, we enable more investigations to understand and verify results at a detailed and complete level.

Most of the samples in our data set are collected from \texttt{kernelmode\-.info}, \texttt{VirusTo\-tal}, and a few are from \texttt{virusshare.com}, \texttt{contagi\-odump.com} and Malpedia \cite{CybIN}. To ensure that we had no benign samples we required each sample to be detected by at least 13 anti-malware vendors and to ensure consistency of the samples belonging to a given malware family we required for each sample that at least two vendors label it within the same malware family. On average each sample had 51 anti-malware vendors report it as malicious and a median of 53.

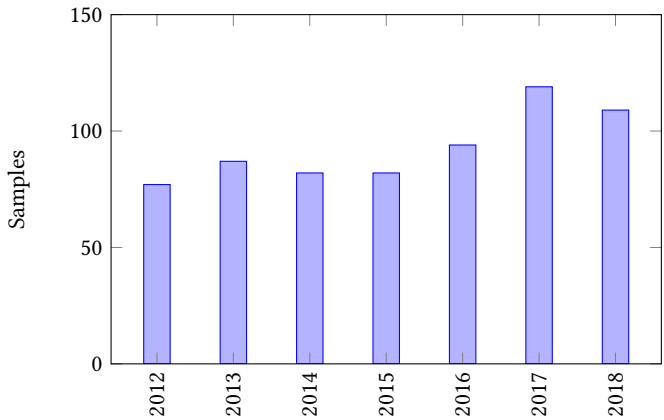
\begin{figure}[]
\centering
\hspace*{-0.5cm} 
\begin{tikzpicture}
\begin{axis}[
    width=0.5\textwidth,
    height=0.35\textwidth,
    ymin=0, ymax=150,
    xtick = {2012,2013, ..., 2018},
    area style,
    xlabel=Year sample was first submitted to VirusTotal.,
    x label style={at={(axis description cs:0.5,-0.2)},anchor=north},
    ylabel=Samples,
    x tick label style={rotate=90, /pgf/number format/1000 sep=}
    ]
\addplot+[ybar] plot coordinates {
(2012,77)
(2013,87)
(2014,82)
(2015,82)
(2016,94)
(2017,119)
(2018,109)
};
\end{axis}
\end{tikzpicture}
\caption{Yearly-distribution of samples in our data set.}
\label{fig:yearlydistributionofsamples}
\end{figure}

\begin{figure}
\centering
\begin{tikzpicture}
\begin{axis}[
    width=0.5\textwidth,
    height=0.35\textwidth,
    ymin=0, ymax=45,
    xtick = {2012,2013, ..., 2018},
    area style,
    xlabel=Year,
    x label style={at={(axis description cs:0.5,-0.2)},anchor=north},
    ylabel=Families,
    x tick label style={rotate=90, /pgf/number format/1000 sep=}
    ]
\addplot+[ybar] plot coordinates {
(2012, 19)
(2013, 31)
(2014, 33)
(2015, 31)
(2016, 27)
(2017, 27)
(2018, 18)};
\end{axis}
\end{tikzpicture}
\caption{Yearly-distribution of families.}
\label{fig:YearlyFistributionOfFamilies}
\end{figure}
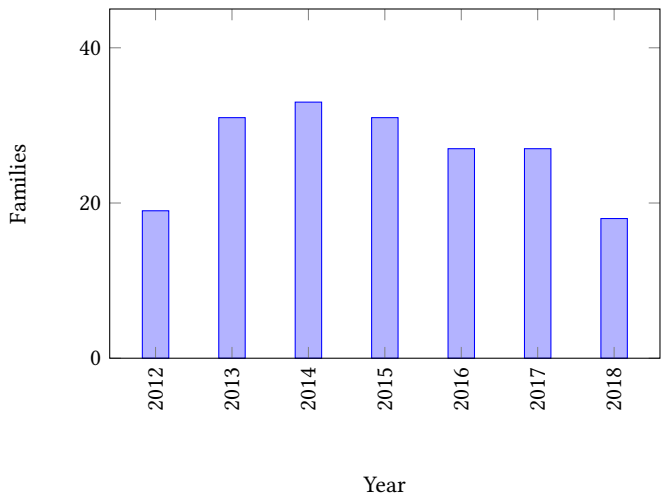

Our data set contains 65 different malware families with ten samples from each family, making a total of 650 samples. Table \ref{tab:malware_samples_large_experiment} shows the specific families and the years their samples were first discovered, and Figure \ref{fig:yearlydistributionofsamples} shows the yearly-distribution of when samples were first discovered. Our data set has samples spanning 2012-2018 and Figure \ref{fig:YearlyFistributionOfFamilies} shows the number of families represented each year. We have an average of 22 families represented each year, with a maximum of 33 families in 2014 and a minimum of 19 families in 2012.

\begin{table}{}
\hspace*{-0.0cm} 
\centering
\footnotesize
    \begin{tabular}{|l|c|c||c|}
    \hline
         Family & Discovered& Family & Discovered \\
    \hline
Androm & 2012-2014  & Midie & 2013, 2016-2017   \\
Artemis & 2012  & MiniDuke & 2013, 2017   \\
Barys & 2012-2013, 2015  & Mira & 2014, 2016-2017   \\
Bitman & 2015-2016 & Natas & 2017-2018   \\
Buzus & 2013, 2016, 2018  & Neshta & 2012-2014   \\
CTBLocker & 2015-2016  & Neshuta & 2012-2014   \\
Cerber & 2015-2017  & Ngrbot & 2012, 2014-2018   \\
CoinMiner & 2013-2015, 2017  & Nimnul & 2013-2017   \\
CosmicDuke & 2013-2014  & Nitol & 2013, 2015   \\
Crowti & 2014-2016  & Nymaim & 2013, 2015-2018   \\
Cryptlock & 2013-2014, 2016  & Otwycal & 2017-2018   \\
Cutwail & 2014-2016, 2018  & Padodor & 2017-2018   \\
DealPly & 2013-2014 & Parite & 2013, 2015, 2017   \\
Dorkbot & 2012, 2013, 2015, 2018  & Pony & 2014, 2018   \\
Dridex & 2014-2016  & Pronny & 2018   \\
Eldorado & 2013-2014  & Ramnit & 2012, 2014-2017   \\
Emotet & 2012, 2014-2017  & Razy & 2013-2014   \\
Fareit & 2013, 2015-2017  & Renos & 2014, 2018   \\
Flood & 2013, 2015-2017  & Rovnix & 2014, 2016   \\
Fujacks & 2014-2015, 2018  & Sality & 2014, 2016-2017   \\
Fynloski & 2012-2013, 2015-2016, 2018  & Shifu & 2015, 2017   \\
Gamarue & 2012, 2017   & Simda & 2012, 2014-2015   \\
Gootkit & 2014-2015   & Symmi & 2012-2014   \\
Kasidet & 2015, 2018   & TeslaCrypt & 2015-2016   \\
Kazy & 2013-2014, 2017   & TinyBanker & 2012, 2017   \\
Kovter & 2013-2016   & Urausy & 2012   \\
Kraddare & 2012-2014   & Ursnif & 2015-2016, 2018   \\
Kryptik & 2012-2015, 2017   & VBKrypt & 2013-2014, 2016-2017   \\
Madangel & 2013, 2015, 2017   & Vawtrak & 2013, 2015-2016, 2018   \\
Madi & 2012   & Wannacry & 2017-2018   \\
Mamba & 2018   & Waski & 2013-2015, 2017   \\
Mazam & 2016   & Zbot & 2012-2014   \\
Vilsel & 2014, 2016-2017   & & \\
\hline
\multicolumn{4}{|l|}{\textbf{Total families: 65}} \\
\multicolumn{4}{|l|}{\textbf{Total samples: 650}} \\
\hline
    \end{tabular}

    \caption{The malware families that we used for our study. We used 10 samples from each family.}
    \label{tab:malware_samples_large_experiment}
\end{table}

\subsection{Research questions}
The goal of our study is to characterise malware propagation at large and to do this, we propose four high level research questions. 

\begin{enumerate}
    \item \textit{Is system-wide malware propagation prevalent? }
    
    The goal of this research question is to give quantitative measurements of the basic characteristics of malware propagation at large. To answer this, we gather statistics on the number of processes and execution waves in each malware sample and also the depths and width of the SPG.   

    \item \textit{Are the propagation strategies used in the wild diverse?}
    
    The goal of this research question is to give insights about the individual propagation strategies used by the malware samples. To answer this, we investigate (1) which processes are the targets of malware propagation, (2) which specific techniques are used for malware propagation and (3) the proportion of droppers versus code injections.
    
    \item \textit{Are there clear relations between malicious behaviour and system-wide malware propagation?}
    
    The goal of this research question is to identify how malware performs its malicious behaviours, e.g. achieves persistence, escalates process-level privileges or connects back to its C\&C server, relative to its system-wide propagation strategy. To answer this question, we investigate where in the SPG malware: (1) uses dynamically generated code, (2) executes most of its code and (3) performs sensitive API calls. 

    \item \textit{Has system-wide malware executions changed consistently towards one direction and are there any clear inter-family SPG consistency?}
    
    The goal of this research question is to measure if malware propagation has changed over time and if samples from the same family propagate similarly. To answer this question, we gather statistics based on grouping our data set into the years the samples were first observed and also do family-oriented statistics.

\end{enumerate}

\subsection{Experimental set up}
\label{sec:Chapter6ExperimentalSetUp}
We conduct all of our experiments on a 4-core Intel-7 CPU with 4.2 GHz and a Windows 7, 32-bit guest architecture. The guest is in a closed network and connected to another virtual machine that performs network simulation using InetSim\cite{InetSim}. As such, malware samples that connect back to some C\&C server will be able to resolve DNS names, connect to every IP and also receive content. However, the content itself is the default data provided by INetSim, which will hinder malware samples from getting C\&C-specific data, e.g. bot commands. We set a recording time of 25 seconds and a loft of 120 minutes for replaying. Due to the resources required to run this type of experiment, we analysed each sample once. 

The applications on the guest machine were executed with a local admin account and with User Account Control (UAC) enabled. We perform no user stimulation during the analysis, and no applications were running in the guest system apart from the generic Windows processes. To start execution of a given sample, we mount it as an \texttt{.iso} file in QEMU and then send keystrokes to QEMU to execute command-line commands in the guest's terminal for opening the mounted \texttt{.iso} and starting the sample. 

\subsection{Generating multi-process signatures}
\label{sec:chapter6GenMultiSigs}
The goal of our study is to give a precise view of system-wide propagation and we need to ensure we distinguish malware samples that use different multi-process propagation techniques, while simultaneously identifying samples that rely on the same technique. To this end, we use the output of Minerva to highlight multi-process propagation and then manually refine this information into easily recognisable signatures. We use the precise API capture of Minerva, the system-wide malware execution trace and identification of intrinsic aspects in multi-process transitions  to speed up the reverse engineering process significantly. We started our study with no signatures and then incrementally created signatures for each analysed sample.

\subsection{False positive and false negative elimination}
\label{sec:chapter7FalsePositiveElimination}
Minerva may over-taint the system and include benign code in the malware execution trace. To elegantly solve this issue, we introduce a heuristic that identifies over-tainting in order to eliminate false positives.

To identify over-tainting, we first check for each sample with multi-process execution if any of our generated signatures recognise the multi-process transitions. If there are some processes with malicious execution but no matching signature, then we consider this process a potential case of over-tainting. We determine this by matching the tainted code with code that was present in the system before the malware execution. The idea is that if the code execution is due to over-tainting, it must be part of the non-malware code and, therefore, part of the original code as it was before the malware execution. We verify this in practice by matching API calls observed by Minerva and the API calls performed by the potentially benign code. If we find more than 99.00\% of the API calls from the tainted code correspond to API calls in the existing Windows code or if the tainted code has no API calls in the process at all then we declare the process a false-positive. Otherwise, we consider it a true positive and reverse engineer the code in order to create a signature. The reason we maintain a 99\% threshold is to allow for incompleteness in our ability to extract API calls statically from the Windows code and Minerva does this as part of our static analysis post dynamic analysis. So far we have not found any false positives from this approach.

Finally, PANDA itself may fail in some instances and only replay little of the recorded execution. To deal with samples that did not execute properly, we remove all samples with less than 25 different instructions executed on the system. 

All of the statistics we report in this paper are post false-positive elimination, including the input data set described in \ref{sec:ChapterEmpiricalStudy_data_collection}. As such, we have manually verified multi-process execution for all samples in our data set as a result of our signature creation. The numbers we report are, therefore, a strict lower-bound, but we may miss out on some aspects of the malware capabilities due to the nature of dynamic analysis. Although a strict lower-bound, we consider it a precise estimate, and often significantly more accurate in comparison to other work, as justified by the work described in \cite{2019arXiv190809204K}.

\section{Experimental results}
In this section, we present the results of our study in terms of the four research questions described above. This section focuses on a quantitative presentation and then in Section \ref{chapter6_discussion_of_results}, we give an interpretation and qualitative discussion of the results. 

\subsection{Quantitative measurements}
We now present results related to our first research question ``\textit{Is system-wide malware propagation prevalent?}'' by gathering statistics about the core attributes of the SPGs in our entire data set. 

\begin{figure}
\centering
\hspace*{-0.5cm} 
\begin{minipage}{0.5\textwidth}
\centering
\begin{adjustbox}{scale=0.6}
\begin{tikzpicture}
	%\node (wa) [wa] {wave1};

	\node (300) [wave2] {\texttt{sample.exe}\\\textbf{$318$}, 1 wave};
	\path (300)+(0.0, -2.5) node (318) [wave2] {\texttt{sample.exe}\\\textbf{$2f8$}, 1 wave};
	\path (318)+(0.0, -2.5) node (2fc) [wave2] {\texttt{ulowu.exe}\\\textbf{$2f0$}, 2 waves};
	\path (318)+(5.5, +0.0) node (2f0) [wave2] {\texttt{cmd.exe}\\\textbf{$6cc$}, 1 wave};
	\path (2fc)+(0.0, -2.5) node (14c) [wave2] {\texttt{ulowu.exe}\\\textbf{$150$}, 1 waves};

	\path (14c)+(-6.0, +0.0) node (43c) [wave2] {\texttt{explorer.exe}\\\textbf{$43c$}, 1 wave};
	\path (14c)+(5.5, 0.0) node (76c) [wave2] {\texttt{conhost.exe}\\\textbf{$76c$}, 1 wave};

	\path (14c)+(5.5, -2.5) node (73c) [wave2] {\texttt{cmd.exe}\\\textbf{$73c$}, 1 wave};
    \path (14c)+(-6.0, -2.5) node (458) [wave2] {\texttt{dwm.exe}\\\textbf{$458$}, 1 wave};
    
    \path (14c)+(5.5, -5.0) node (6cc) [wave2] {\texttt{conhost.exe}\\\textbf{$14c$}, 1 wave};
    \path (14c)+(+0.0, -5.0) node (50c) [wave2] {\texttt{taskhost.exe}\\\textbf{$50c$}, 1 wave};

	\fill (300.west)+(-0.7,0) circle (0.1cm);	
	\path (300.west)+(-0.7,0) node (EOP) {};
	\path [draw, ->] (EOP) -- node [above] {} (300);	
	
	\path [draw, ->] (318.east) -- node [below] {} (2f0.west); 
	
	\path [draw, ->] (300.south) -- node [above] {} (318.north);
	\path [draw, ->] (318.south) -- node [above] {} (2fc.north);
	\path [draw, ->] (2fc.south) -- node [above] {} (14c.north);
	
	\path [draw, ->] (43c.10) -- node [above] {} (14c.170);
	\path [draw, ->] (14c.190) -- node [above] {} (43c.350);
    
    \path [draw, ->] (14c.205) -- node [below] {} (458.east);

    \path [draw, ->] (14c.south) -- node [below] {} (50c.north);

    \path [draw, ->] (14c.315) -- node [below] {} (6cc.west);
    \path [draw, ->] (14c.335) -- node [below] {} (73c.west);
    \path [draw, ->] (14c.east) -- node [below] {} (76c.west);

    \path [draw, ->] (14c.25) -- node [below] {} (2f0.205);
\end{tikzpicture}
\end{adjustbox}
\end{minipage}
\captionsetup{justification=centering,margin=1cm}
\caption{The system-wide propagation graph of Natas sample. Only processes are shown, and each process box contains the \texttt{PID} and the number of waves for the given process.}
\label{fig:NatasMalware}
\end{figure}
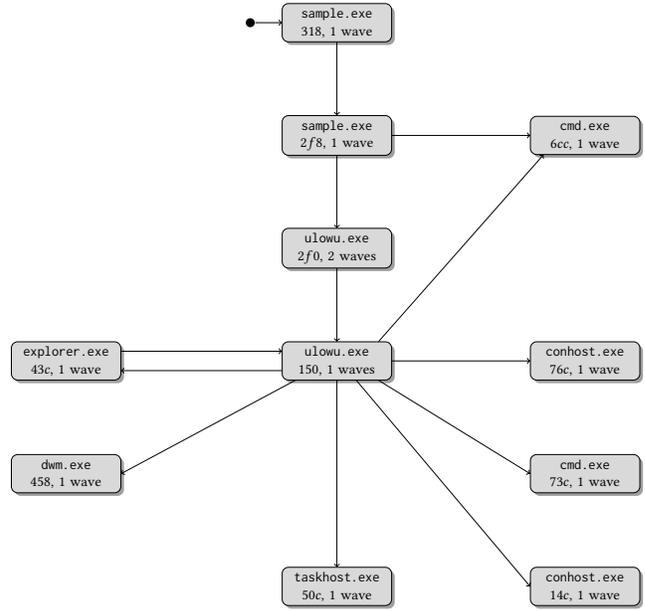

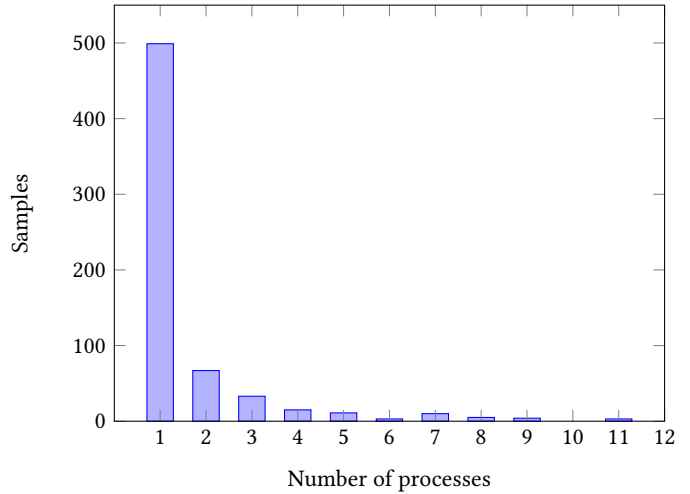
\begin{figure}
\centering
\begin{tikzpicture}
\begin{axis}[
    width=0.5\textwidth,
    height=0.4\textwidth,
    ymin=0, ymax=550,
    xtick = {1,2, ..., 15},
    area style,
    xlabel=Number of processes,
    ylabel=Samples
    ]
\addplot+[ybar] plot coordinates { 
(1, 499)
(2, 67)
(3, 33)
(4, 15)
(5, 11)
(6, 3)
(7, 10)
(8, 5)
(9, 4)
(11, 3)};
\end{axis}
\end{tikzpicture}
\caption{Number of processes involved in malware executions of all samples.}
\label{fig:numberofprocessesexecuted}
\end{figure}

The first statistic we gather is the number of processes involved in each malware execution, and Figure \ref{fig:numberofprocessesexecuted} shows the results. In total, 151 of the 650 samples exhibited multi-process propagation and 67 of these samples split into two processes. This means 23.23\% of all samples use multi-process propagation and 55.6\% of malware that performs multi-process propagation do so in three or more processes. The 151 samples are from 40 different families corresponding to 62\% of all families, and Table \ref{tab:SamplesInFamiliesWithMultiProc} shows the complete list of families and their respective number of multi-process samples. Madangel is the only family where all samples use multi-process propagation, and we observed four families where all but one sample deployed multi-process execution, namely Emotet, Razy, Natas and TinyBanker. The maximum number of processes we observed in a single malware execution is eleven, which occurred in three samples from the Natas family and Figure \ref{fig:NatasMalware} shows the SPG of one of these samples\footnote{\texttt{SHA-256 sum bd037ee28fc97f59525f384136fc067dded691230597384f04b10efe360055d1}}. The malware initially transitions through three processes via files that it dropped on the system and then at a process-depth of four injects code into every process on the system for which it has permissions. 

\begin{table}[]
    \centering

    \footnotesize
    \begin{tabular}{|lc|cc|cc|cc|}
    \hline
         Family & $|\mathcal{M}|$ & Family & $|\mathcal{M}|$ & Family & $|\mathcal{M}|$ & Family & $|\mathcal{M}|$\\
        \hline
Madangel & 10 &  CoinMiner & 5 & Ursnif & 3 &  Fynloski & 2\\
Natas & 9 &  Nitol & 5 & Dridex & 3 &  Cerber & 1\\
Emotet & 9 &  Midie & 4 & CTBLocker & 3 &  Barys & 1\\
TinyBanker & 9 &  Urausy & 4 & Kryptik & 3 &  Bitman & 1\\
Razy & 9 &  Vawtrak & 4 & VBKrypt & 2 &  Fareit & 1\\
Gamarue & 7 &  Dorkbot & 3 & Androm & 2 &  Waski & 1\\
Ramnit & 7 &  Zbot & 3 & Buzus & 2 &  TeslaCrypt & 1\\
Mira & 7 &  Kasidet & 3 & Cutwail & 2 &  Eldorado & 1\\
Sality & 6 &  Kovter & 3 & Crowti & 2 &  Shifu & 1\\
Nimnul & 6 &  Symmi & 3 & Nymaim & 2 &  Ngrbot & 1\\

         \hline
    \end{tabular}
    \caption{Families with multi-process propagation. $|\mathcal{M}|$ denotes the number of multi-process samples in the given family.}
    \label{tab:SamplesInFamiliesWithMultiProc}
\end{table}

The process-depths exhibited by the samples are shown in Table \ref{tab:process_depths} and, in total, 91.7\% of the samples have a process-depth of either one or two. This is more samples than the number of samples with one or two process executions (87\%) confirming that some malware split execution rather than incrementally build a chain of multi-process executions. We found a maximum process-depth of five, and this occurred in seventeen samples. The complete list of process-depths for all 40 families with multi-process propagation is shown in Table \ref{tab:famSPGProcDepths}. We find that 17 families have multiple samples with multi-process execution that do not exhibit the same process-depth, and, on the other hand, we find that in 12 families with multiple multi-process samples all samples with multi-process execution within their respective family exhibit the same process-depth. 

The SPG-widths of all the samples are shown in Table \ref{tab:wave_breath_statistics}. In total, 597 of the samples have an SPG-width of one, which means the vast majority of samples sequentially propagate through processes. The maximum SPG-width we found was seven, and this occurred in five malware samples, three of which belong to the Natas family, one from the Shifu family and one from the Ramnit family. We found a total of 20 families with at least one sample that has SPG-width larger than one, and the full list of these families are shown in Table \ref{tab:famSPGwidths} alongside the number of samples in each family that deploys a given SPG-width. 

\begin{table}[]
    \centering
    \footnotesize
    \begin{tabular}{|c|c|c|c|c|c|c|}
    \hline
         Process-depth & 1 & 2 & 3 & 4 & 5 \\
        \hline
         Samples & 499 & 97 & 23 & 14 & 17 \\
         \hline
    \end{tabular}
    \caption{Process-depths of all samples.}
    \label{tab:process_depths}
\end{table}

\begin{table}[]
    \centering
    \scriptsize
    \begin{tabular}{c|c|c|c|c|c|c|c|c|c|c|c}
        \hline
        Process-depth & 1 & 2 & 3 & 4 & 5 & & 1 & 2 & 3 & 4 & 5 \\
        \hline
Androm & 8  & 1  & 1  &  & & Midie & 6  & 3  &  &  & 1 \\
Barys & 9  &  & 1  &  & & Mira & 3  & 7  &  &  & \\
Bitman & 9  & 1  &  &  & & Natas & 1  & 4  &  &  & 5 \\
Buzus & 8  &  &  & 1  & 1 & Ngrbot & 9  & 1  &  &  & \\
CTBLocker & 7  & 3  &  &  & & Nimnul & 4  & 3  & 1  & 2  & \\
Cerber & 9  & 1  &  &  & & Nitol & 5  & 5  &  &  & \\
CoinMiner & 5  & 5  &  &  & & Nymaim & 8  & 2  &  &  & \\
Crowti & 8  & 1  & 1  &  & & Ramnit & 3  & 3  & 2  & 1  & 1 \\
Cutwail & 8  & 2  &  &  & & Razy & 1  & 9  &  &  & \\
Dorkbot & 7  & 1  & 2  &  & & Sality & 4  & 6  &  &  & \\
Dridex & 7  & 2  & 1  &  & & Shifu & 9  &  & 1  &  & \\
Eldorado & 9  & 1  &  &  & & Symmi & 7  & 1  & 1  & 1  & \\
Emotet & 1  & 1  & 1  & 5  & 2 & TeslaCrypt & 9  & 1  &  &  & \\
Fareit & 9  & 1  &  &  & & TinyBanker & 1  &  & 7  & 1  & 1 \\
Fynloski & 8  & 2  &  &  & & Urausy & 6  & 1  &  &  & 3 \\
Gamarue & 3  & 7  &  &  & & Ursnif & 7  & 1  &  & 2  & \\
Kasidet & 7  & 3  &  &  & & VBKrypt & 8  &  &  &  & 2 \\
Kovter & 7  & 1  & 1  & 1  & & Vawtrak & 6  & 2  & 2  &  & \\
Kryptik & 7  & 3  &  &  & & Waski & 9  & 1  &  &  & \\
Madangel &  & 10  &  &  & & Zbot & 7  & 1  & 1  &  & 1 \\
    \end{tabular}
    \caption{Families with at least one sample that has process-depth higher than one. The columns indicate a given process-depth and the rows display the number of samples in a given family with the given depth.}
    \label{tab:famSPGProcDepths}
\end{table}

\begin{table}[]
    \centering
    \small
    \begin{tabular}{|c|c|c|c|c|c|c|c|}
        \hline
        SPG-width & 1 & 2 & 4 & 4 & 5 & 6 & 7 \\
        \hline
        Samples &     597 & 24 & 4 & 6 & 9 & 5 & 5 \\
        \hline
    \end{tabular}
    \caption{SPG-width of all samples.}
    \label{tab:wave_breath_statistics}
\end{table}

\begin{table}[]
    \centering
    \scriptsize
    \begin{tabular}{c|c|c|c|c|c|c|c}
        \hline
        SPG-widths & 1 & 2 & 3 & 4 & 5 & 6 & 7 \\
        \hline
        Androm & 9 & 1 & & & & & \\
CTBLocker & 7 & 3 & & & & &\\
Cerber & 9 & 1  & & & & &\\
CoinMiner & 9 & & 1 & & & &\\
Cutwail & 8 & 2 & & & & &\\
Dorkbot & 8 & & & & 2 & & \\
Emotet & 3  &4  & &2  &1  & &\\
Kovter & 9  &1  & & & & &\\
Madangel & 8  &2  & & & & &\\
Midie & 9  &1  & & & & &\\
Natas & 5  & & & &2  & &3  \\
Nitol & 5  &5  & & & & &\\
Nymaim & 8  &2  & & & & &\\
Ramnit & 5  &1  &3  & & & &1 \\
Razy & 9  &1  & & & & &\\
Sality & 4  & & & &1  &5  &\\
Shifu & 9  & & & & & &1  \\
TinyBanker & 8  & & &2  & & &\\
VBKrypt & 8  & & &2  & & &\\
Vawtrak & 7  & & & &3  & &\\
    \end{tabular}
    \caption{Families with at least one sample that has SPG-width higher than one. The columns indicate a given SPG-width and the rows display the number of samples in a given family that deploys the given SPG-width.}
    \label{tab:famSPGwidths}
\end{table}

The distribution of execution waves is shown in Table \ref{tab:execution_wave_distribution}. In total, 393 of 650 samples have multiple execution waves, which correspond to 60\% of the samples, and we found thirteen samples that deploy more than 25 execution waves. The three highest counts of execution waves are 54, 354 and 5389, which came from samples in the Pronny, Artemis and Urausy malware families, respectively. In figure \ref{fig:PerFamilyExecutionWaves} we show the average number and standard deviation of execution waves per family in all families except Pronny, Artemis and Urausy, who have an average of 43, 37 and 545 execution waves, respectively, and, a standard deviation of 2, 52 and 807. We observe seven families that have no execution waves, meaning they do not deploy any dynamically generated code, and the average number of execution waves is slightly less than 15 if we discard the three families mentioned above. Additionally, the standard deviation is reasonably consistent across the majority of families, however, certain families show a considerable variation and the following families: Urausy, Artemis, Shifu, Buzus, Mira, Androm, Ursnif, Symmi and VBkrypt each have a standard deviation that is larger than their respective average number of execution waves. 

\begin{table}{}
    \centering
    \hspace*{-0.6cm} 
    \footnotesize
    \begin{tabular}{|l|c|c|c|c|c|c|c|}
        \hline
         \textbf{Execution waves} & 1 & 2 & 3 & 4 & 5 & 6 & 7  \\
         Samples         & 257 & 123 & 97 & 41 & 17 & 31 & 7  \\
         \hline \hline
         \textbf{Execution waves} & 8 & 9 & 10 & 11-15 & 16-25 & $>$25 & \\
         Samples & 12 & 6 & 7 & 26 & 13 & 13 & \\
         \hline
    \end{tabular}
    \centering
    \caption{The number of execution waves of all samples.}
    \label{tab:execution_wave_distribution}
\end{table}

\subsection{Propagation intrinsics}
We now present results related to our second research question ``\textit{Are the propagation strategies used in the wild diverse?}''. To do this, we identify the processes that are targets of multi-process propagation, identify the API routines used by malware to propagate through the system, discuss detailed case studies of interesting malware samples, and also measure the proportion of code injections versus propagations via dropped files.

\subsubsection{Target processes}
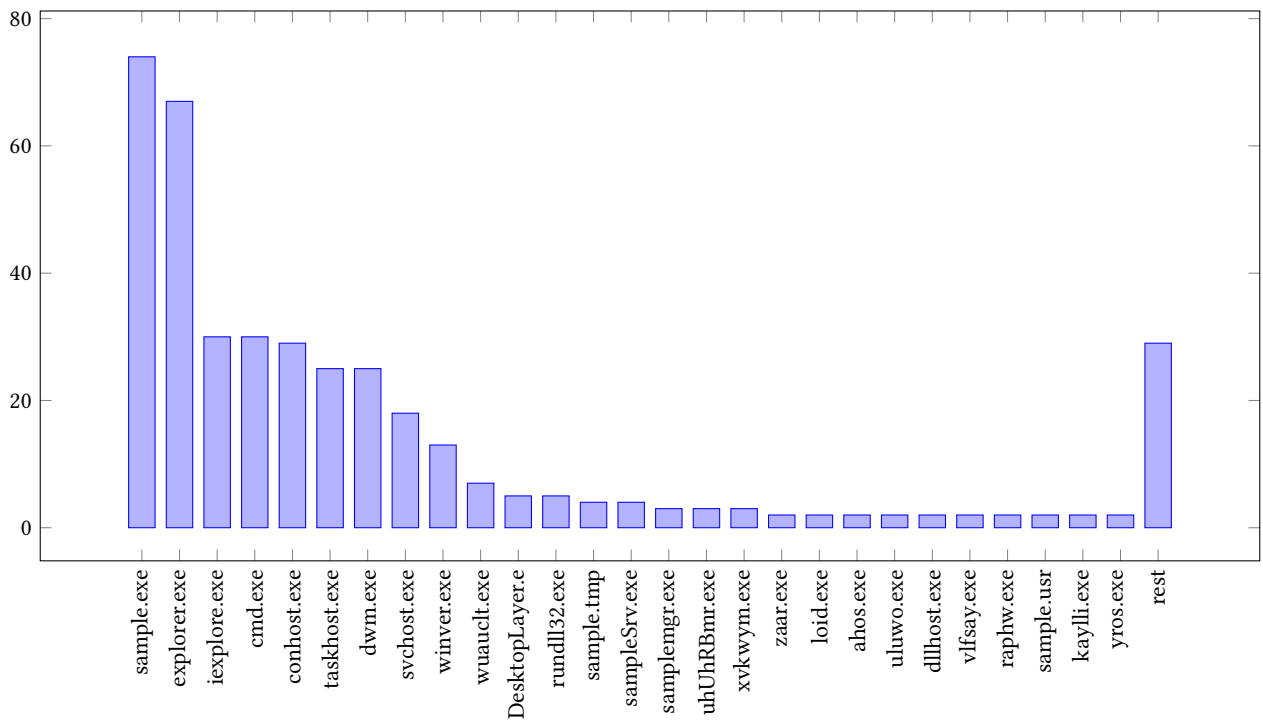
\begin{figure*}
\centering
\begin{tikzpicture}
\begin{axis}[
    width=1.0\textwidth,
    height=0.5\textwidth,
    symbolic x coords={sample.exe, explorer.exe, iexplore.exe, cmd.exe, conhost.exe, taskhost.exe, dwm.exe, svchost.exe, winver.exe, wuauclt.exe, DesktopLayer.e, rundll32.exe, sample.tmp, sampleSrv.exe, samplemgr.exe, uhUhRBmr.exe, xvkwym.exe, zaar.exe, loid.exe, ahos.exe, uluwo.exe, dllhost.exe, vlfsay.exe, raphw.exe, sample.usr, kaylli.exe, yros.exe, rest},
    x tick label style={rotate=90},
        area style,
    xtick=data]
    \addplot+[ybar] plot coordinates {
(sample.exe, 74)
(explorer.exe, 67)
(iexplore.exe, 30)
(cmd.exe, 30)
(conhost.exe, 29)
(taskhost.exe, 25)
(dwm.exe, 25)
(svchost.exe, 18)
(winver.exe, 13)
(wuauclt.exe, 7)
(DesktopLayer.e, 5)
(rundll32.exe, 5)
(sample.tmp, 4)
(sampleSrv.exe, 4)
(samplemgr.exe, 3)
(uhUhRBmr.exe, 3)
(xvkwym.exe, 3)
(zaar.exe, 2)
(loid.exe, 2)
(ahos.exe, 2)
(uluwo.exe, 2)
(dllhost.exe, 2)
(vlfsay.exe, 2)
(raphw.exe, 2)
(sample.usr, 2)
(kaylli.exe, 2)
(yros.exe, 2)
(rest, 29)
    };
\end{axis}
\end{tikzpicture}
\caption{The processes that are the most popular targets for multi-process propagation.}
\label{fig:SpecificProcessTargets}
\end{figure*}

Table \ref{fig:SpecificProcessTargets} shows the names of the most targeted processes and the number of times they were used for multi-process propagation. The most popular target is malware starting another instance of itself, in this case \texttt{sample.exe}. The second most popular target is \texttt{explorer.exe} and following this are \texttt{iexplore.exe}, \texttt{cmd.exe} and \texttt{conhost.exe}, which are all part of the Windows OS. In fact, the following eight Windows applications \texttt{explorer.exe}, \texttt{iexplore.exe}, \texttt{cmd.exe}, \texttt{conhost.exe}, \texttt{task\-host.exe}, \texttt{dwm.exe} \texttt{svchost.exe}, and \texttt{winver.exe} make up 241 targets out of 394 multi-process propagations, corresponding to roughly 61\%. Furthermore, if we discard the multi-process executions via \texttt{sample.exe} these eight applications make up 75\% of the victims. 

In total, we found 244 multi-process executions targeted processes with names from the Windows OS, and 150 multi-process executions with proprietary names where \texttt{sample.exe} takes up 74 of these. As such, there are only 76 cases where the malware execute via programs with non-standard names, corresponding to 19\% of the total multi-process propagations.

In Table \ref{tab:FamilyTargetCounts}, we show the per-family process targets as well as the number of samples in each family that execute in the given target. The table also shows the size of the group of samples that share most target processes and the size of this group over the total samples in the family.  Finally, the table shows the number of samples in the group just mentioned multiplied by the number of targets they share over the total number of propagations in the given family. We find a total of twenty families where all samples in the given family that deploys multi-process propagation also share the largest intersection of target processes. In seventeen of these families this intersection makes up the entire set of process targets from the given family, and in the remaining three families, namely Sality, TinyBanker and Madangel, some samples also propagate to more targets. In the twenty families, however, only eleven have more than one sample that does multi-process propagation, meaning that in a total of six families with multi-process propagation all samples that propagate into multiple processes share the exact same targets.  Sality is particularly interesting in that all six samples that do multi-process propagation execute within the same five processes, whereas four of these samples also execute in \texttt{rundll.exe} and one also executes in \texttt{sample.tmp}. Gamarue is another interesting case, where all seven samples that do multi-process propagation inject into the same process, namely \texttt{wuauclt.exe}. 

\subsubsection{Signature diversity}
\label{sec:signature_diversity}
Malware use a variety of Windows APIs to propagate through the system, and we now present these by way of the process propagation signatures we created in our analysis of all the multi-process propagations. The signatures we have created are API-specific, e.g. we consider \texttt{CreateFileW} and \texttt{CreateFileA} to be different API functions. The benefit of this is to have detailed statistics, from which we can indeed build abstractions, and the drawback is that some signatures can have minor differences to other signatures, e.g. whether the malware uses Unicode or ASCII. 

We found a total of 417 injections, which is more than the 394 processes that are results of multi-process execution. The reason for this is that multiple processes may inject into the same target process, as shown in the previously discussed Natas sample in Figure \ref{fig:NatasMalware} where multiple processes inject into the same \texttt{cmd.exe} process. The complete set of signatures we observed is shown in Table \ref{tab:SignatureDistribution}, which is sorted by the number of times they occur. In total, we discovered 33 different signatures and the most common way for malware to inject into other processes is using the traditional code injection technique with API calls to \texttt{OpenProcess}, \texttt{VirtualAllocEx}, \texttt{WriteProcessMemory}, \texttt{CreateRemoteThread}.  This approach is used in 174 of 417 cases, corresponding to 41.7\%. In contrast, we found six signatures that are only used once, and they rely mainly on a subset of the APIs from other signatures.  

\newcolumntype{M}{>{\begin{varwidth}{5cm}}l<{\end{varwidth}}}

In general, when malware propagate through the system, it can do so by either launching a new process or migrating to an existing one. To estimate the proportion of these two we count the number of signatures that use \texttt{OpenProcess}, \texttt{CreateProcess} and \texttt{ShellExecute} to access the victim process, which is 211, 213 and 15, respectively.  As such, there is a an almost even distribution between cases that open an existing process (\texttt{OpenProcess}) and launching a new process (\texttt{CreateProcess} and \texttt{ShellExecute}). 

In Table \ref{tab:FamiliesInjectionSignatures}, we show the per-family signature usage, the number of injections that rely on a given signature and the number of samples that use the given signature. We find 17 families that rely on a single signature and 23 families that deploy more than one signature. Sality is an interesting family in that it has a total of 41 process-propagations amongst six samples that all rely on the same signature to propagate. We observe similar trends in Dorkbot, Emotet, Natas, Ramnit, TinyBanker, VBKrypt and Vawtrak where samples extensively reuse the same signatures. We observe an opposite trend in families like Buzus, Dridex, Kovter, Kryptik, Fynloski and Ursnif that each has multiple samples performing multi-process propagation without any overlap in propagation technique. Buzus is particularly interesting in this context, where we observe two samples that combined use seven different injection techniques, and each of the signatures is only used once. In the following section, we will go into details with interesting injection techniques, and then in Section \ref{chapter6:InterFamilyConsistencies}, we will present further quantitative analysis of inter-family consistency in terms of propagation signatures.
\begin{table}
\scriptsize
\centering  % better than "\begin{center}" and "\end{center}"
\begin{tabular}{|l|c|c|l|}
\hline
        \hline
        Family & $|\mathcal{M}|$ & $\Sigma$ & (signature ID, injection count, sample count), ... \\
        \hline
Androm & 2 & 3 &(2, 2, 1) (17, 1, 1) (22, 1, 1) \\
Barys & 1 & 1 &(4, 2, 1) \\
Bitman & 1 & 1 &(2, 1, 1) \\
Buzus & 2 & 7 &(1, 1, 1) (3, 1, 1) (5, 1, 1) (10, 1, 1) (17, 1, 1) (28, 1, 1) (29, 1, 1) \\
CTBLocker & 3 & 1 &(2, 6, 3) \\
Cerber & 1 & 2 &(8, 1, 1) (11, 1, 1) \\
CoinMiner & 5 & 5 &(8, 2, 2) (11, 2, 1) (25, 1, 1) (26, 1, 1) (27, 1, 1) \\
Crowti & 2 & 1 &(4, 3, 2) \\
Cutwail & 2 & 1 &(2, 4, 2) \\
Dorkbot & 3 & 3 &(1, 12, 2) (2, 2, 2) (15, 1, 1) \\
Dridex & 3 & 3 &(1, 1, 1) (4, 2, 1) (33, 1, 1) \\
Eldorado & 1 & 1 &(7, 1, 1) \\
Emotet & 9 & 6 &(1, 18, 3) (2, 10, 6) (3, 2, 2) (4, 10, 5) (5, 2, 2) (10, 1, 1) \\
Fareit & 1 & 1 &(3, 1, 1) \\
Fynloski & 2 & 2 &(2, 1, 1) (10, 1, 1) \\
Gamarue & 7 & 1 &(9, 7, 7) \\
Kasidet & 3 & 1 &(19, 3, 3) \\
Kovter & 3 & 3 &(3, 1, 1) (9, 3, 1) (14, 3, 1) \\
Kryptik & 3 & 3 &(4, 1, 1) (10, 1, 1) (16, 1, 1) \\
Madangel & 10 & 2 &(1, 2, 2) (6, 10, 10) \\
Midie & 4 & 2 &(2, 2, 1) (6, 6, 3) \\
Mira & 7 & 1 &(12, 7, 7) \\
Natas & 9 & 4 &(1, 34, 5) (3, 14, 9) (10, 5, 5) (23, 3, 3) \\
Ngrbot & 1 & 1 &(1, 1, 1) \\
Nimnul & 6 & 6 &(7, 3, 3) (8, 2, 2) (11, 2, 2) (13, 2, 2) (17, 1, 1) (22, 1, 1) \\
Nitol & 5 & 1 &(2, 10, 5) \\
Nymaim & 2 & 1 &(2, 4, 2) \\
Ramnit & 7 & 7 &(1, 7, 1) (5, 2, 1) (7, 1, 1) (8, 6, 6) (11, 3, 3) (13, 5, 5) (18, 4, 4) \\
Razy & 9 & 3 &(2, 2, 1) (7, 7, 7) (16, 1, 1) \\
Sality & 6 & 1 &(1, 41, 6) \\
Shifu & 1 & 1 &(1, 8, 1) \\
Symmi & 3 & 4 &(2, 2, 2) (4, 2, 1) (17, 1, 1) (22, 1, 1) \\
TeslaCrypt & 1 & 1 &(2, 1, 1) \\
TinyBanker & 9 & 3 &(1, 19, 9) (3, 1, 1) (5, 9, 9) \\
Urausy & 4 & 5 &(14, 3, 3) (15, 2, 2) (20, 3, 3) (21, 3, 3) (24, 2, 2) \\
Ursnif & 3 & 6 &(1, 1, 1) (16, 2, 1) (27, 1, 1) (30, 1, 1) (31, 1, 1) (32, 1, 1) \\
VBKrypt & 2 & 3 &(1, 12, 2) (3, 2, 2) (5, 2, 2) \\
Vawtrak & 4 & 3 &(1, 16, 3) (4, 2, 1) (26, 1, 1) \\
Waski & 1 & 1 &(16, 1, 1) \\
Zbot & 3 & 6 &(1, 1, 1) (2, 1, 1) (15, 3, 2) (16, 1, 1) (17, 1, 1) (25, 1, 1) \\
\hline
\end{tabular}    
    \caption{The per-family list of signatures. $\Sigma$ denotes the number of signatures used by a given family and $|\mathcal{M}|$ denotes the number of samples in a given family that deploys multi-process execution. The rightmost column shows the specific signatures used in each family, the number of injections that uses a given signature and the number of samples in the given family that uses this signature. }
    \label{tab:FamiliesInjectionSignatures}
\end{table} 

\subsubsection{Interesting case studies}

\textbf{Injection via continuously rewriting window procedure.} We observed two injections, IDs \textbf{\#24} and \textbf{\#29}, that use dynamically generated code to hide the API calls involved in their respective injection. Specifically, three malware samples from the Urausy and Buzus families repeatedly overwrite a particular memory region with code that performs a single API call in order to create a chain-like structure that results in a complete injection procedure. The memory region is disguised as a Window procedure and called via the \texttt{CallWindowProcA} function. The effect of this is that the code responsible for the injection is constantly overwritten, remains in memory for a short time, and is never entirely represented in memory but only exposed in multiple temporal pieces. Minerva explicitly observes this by detecting a new wave of dynamically generated code whenever the malware calls \texttt{CallWindowProcA}. This makes it easy to detect that the memory is continuously updated because the execution waves have instructions execute in the same memory region. \\ \\

\textbf{Stealthy injection via ZwCreateUserProcess hooking.} Another interesting code injection is ID \textbf{\#13} which leverages three common API calls \texttt{CreateProcessA}, \texttt{VirtualAllocEx} and \texttt{WritePr\-ocessMemory}, but does not have any call to functions like \texttt{CreateRe\-moteThread} or \texttt{ResumeThread}. In all cases where this signature occur, the target program is \texttt{C:\textbackslash Windows\textbackslash system32\textbackslash svchost.exe} and has \texttt{dwCreationFlags} set to \texttt{\textbf{NULL}}, meaning the program is \textbf{not} started in suspended mode. This is unusual because without a call to \texttt{ResumeThread} or \texttt{CreateRemoteThread} it is, at first sight, unclear how the malware achieves code execution in the victim process. Instead, the malware achieves code execution in the victim process by patching the code of \texttt{CreateProcessA} at run time and then hijacking the Windows code that is in charge of creating a new process. It does this by hooking \texttt{ZwCreateUserProcess} to execute malicious code that will allocate and write memory to the target process, all nested inside the \texttt{CreateProcessA} call. To illustrate this, Figure \ref{fig:code_injection_ZwCreateUserProcess} shows the call graph of samples that rely on this injection technique.
\begin{figure}
\begin{lstlisting}
|\textbf{CreateProcessA(victim\_process)}|
    CreateProcessInternal
    ZwCreateUserProcess
	    |\textbf{\small{Hijack first instruction}}| 
	    |\textbf{\small{of ZwCreateUserProcess}}|
	        |\textbf{LOOP:}|
	            |\textbf{VirtualAlloc}|
	            |\textbf{VirtualProtect}|
	            |\textbf{WriteProcessMemory}|
	...
\end{lstlisting}
\caption{Call graph of code injection that hooks \texttt{ZwCreateUserProcess}. API calls made by the malware are shown in bold.}
\label{fig:code_injection_ZwCreateUserProcess}
\end{figure}

The hook is easy to detect given the output of Minerva from looking at the malware execution trace, shown in Figure \ref{Fig:RamnitHook}. From the export extractor in Minerva we know \texttt{ZwCreateUserProcess} is located at \texttt{0x77c46a98} and given the first instruction is turned into a trampoline to malware code, we can easily deduce that the function is hooked. The benefit of this code injection technique is potential evasion against sandboxes and manual debuggers that only monitor \texttt{CreateProcessA} calls and not nested function calls. As such, if the hooking is not observed, then specific parts of the injection will likely go unnoticed, and the threat may evade analysis.
\\\\
\textbf{Stealthy injection without explicit PID access.} In most cases, the signatures rely on \texttt{CreateProcess(A/W)} or \texttt{OpenProcess} to initiate multi-process propagation. These functions make it is easy for an analyst or sandbox to determine the target process as the output of the functions contains the \texttt{PID} of the process they create/open. Naturally, some injection techniques avoid these functions to hide the multi-process target, and an example of this is signature \textbf{\#20}. Injection \textbf{\#20} retrieves a handle to the target process by calling \texttt{GetShellWindows}, opens the process using \texttt{ZwOpenProcess} and then continues with a familiar pattern of functions to initiate execution. In this way, the \texttt{PID} of the target process is never exposed in any of the functions used in the injection, which adds a level of complexity to identifying the specific propagation procedure. Naturally, Minerva follows the malware execution regardless of knowing the target process \texttt{PID}. 

\begin{figure} 
\centering
\small
\begin{tabular}{| l | l | }
\hline
Address & instruction \\
\hline
0x405996 & jmp \texttt{CreateProcessA(...)}\\
0x77c46a98 & jmp 0x403447 \textbf{\textit{Hook in ZwCreateUserProcess}} \\
0x403447 & push ebp \textbf{\textit{Hooked code}}\\
... & ... \textbf{\textit{Hooked code}}\\
0x4013c0 & lea edi, [ebp-0x54] \textbf{\textit{Return from CreateProcessA}}\\
\hline
\end{tabular}
\caption{The malware execution trace of a code injection that hooks the function \texttt{ZwCreateUserProcess}.}
\label{Fig:RamnitHook}
\end{figure}

\subsubsection{Code injections vs droppers}
An interesting distinction is whether a multi-process propagation is due to code injection or a dropped executable. Although these two approaches are not mutually exclusive (because a sample can drop a file, launch it and then inject code into it) it is interesting to estimate how prevalent each of them is. To give such an estimate we check for each multi-process propagation if the target process corresponds to a file created by the malware earlier in the execution, and if so consider it to be an execution of a dropped executable. We found 52 multi-process executions where the propagation was in a created file and 342 cases where it was not the case. As such, we estimate that 13\% of the multi-process executions in our data set are due to droppers, and the remaining 87\% are due to code injections. We found a total of 14 malware families that had samples propagate via created files and Table \ref{tab:SamplesWithDroppers} lists these together with the number of samples in each family that propagate in this way.

\begin{table}{}
    \centering
    \footnotesize
    \begin{tabular}{|l|c|c|c|c|c|}
        \hline
         Family & $|\mathcal{M}|$ & $|\mathcal{M}_D|$ & Family & $|\mathcal{M}|$ & $|\mathcal{M}_D|$\\
         \hline
Cerber & 1 & 1 & Ramnit & 7 & 6\\
CoinMiner & 5 & 5 & Razy & 9 & 8 \\
Dridex & 3 & 1 & Sality & 6 & 1 \\
Eldorado & 1 & 1 & Urausy & 4 & 2 \\
Kryptik & 3 & 1 & Ursnif & 3 & 2 \\
Natas & 9 & 5 & Waski & 1 & 1 \\
Nimnul & 6 & 5 & Zbot & 3 & 1\\
         \hline
    \end{tabular}
    \centering
    \caption{The families with samples that propagate via dropped files. The table also shows the number of samples with multi-process propagation, denoted $|\mathcal{M}|$, and the unique number of samples in each family that deploys propagation via dropped files, denoted $|\mathcal{M}_D|$.}
    \label{tab:SamplesWithDroppers}
\end{table}

\subsection{Malware activities}
We now move on to the third research question ``\textit{Are there clear relations between malicious behaviour and system-wide malware propagation?}''. This research question aims to answer how malware distributes its activities across its propagation and in this section we, therefore, focus our statistics on the 151 samples that contain multi-process execution and leave the other samples out. We limit ourselves to these because our interest is in highlighting the difference between the initial process and the non-initial processes. The initial process refers to execution at process-depth one, and the non-initial processes refer to execution in processes with a depth greater than one.

\subsubsection{Execution wave distribution}
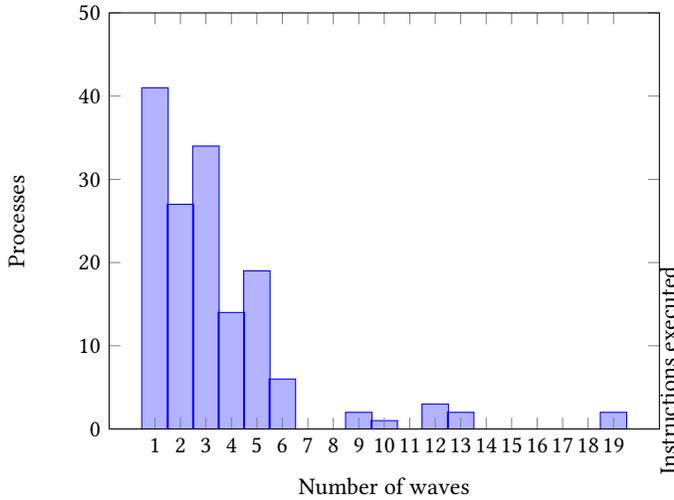
\begin{figure}[h]
\centering
\begin{tikzpicture}
\begin{axis}[
    width=0.5\textwidth,
    height=0.4\textwidth,
    ymin=0, ymax=50,
    xtick = {1,2, ..., 19},
    area style,
    xlabel=Number of waves,
    ylabel=Processes
    ]
\addplot+[ybar] plot coordinates {
(1, 41)
(2, 27)
(3, 34)
(4, 14)
(5, 19)
(6, 6)
(9, 2)
(10, 1)
(12, 3)
(13, 2)
(19, 2)};
\end{axis}
\end{tikzpicture}
\caption{The number of execution waves in the initial process of all multi-process malware.}
\label{fig:NumberOfWavesInMultiProcessMalwareFirstProcess}
\end{figure}

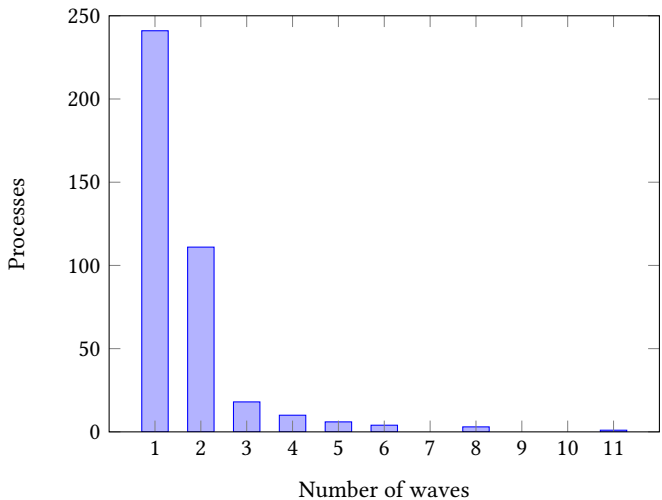
\begin{figure}[h]
\centering
\begin{tikzpicture}
\begin{axis}[
    width=0.5\textwidth,
    height=0.4\textwidth,
    ymin=0, ymax=250,
    xtick = {1,2, ..., 11},
    area style,
    xlabel=Number of waves,
    ylabel=Processes
    ]
\addplot+[ybar] plot coordinates {
(1,241)
(2,111)
(3,18)
(4,10)
(5,6)
(6,4)
(8,3)
(11,1)};
\end{axis}
\end{tikzpicture}
\caption{The number of execution waves in non-initial processes of all multi-process malware.}
\label{fig:NumberOfWavesInMultiProcessMalwarePostFirstProcess}
\end{figure}

The first aspect we analyse is where in the propagation malware deploys dynamically generated code. Specifically, we are interested in knowing if malware continues to use dynamically generated code even after multi-process execution. To do this, we count the number of waves in the initial process and compare it to the number of waves in non-initial processes. Amongst the 151 samples with multi-process execution, we have a total of 545 processes with malware execution, meaning 394 of these are non-initial processes. The distribution of waves in the initial process is shown in Table \ref{fig:NumberOfWavesInMultiProcessMalwareFirstProcess}, and the distribution of waves in non-initial processes is shown in Table \ref{fig:NumberOfWavesInMultiProcessMalwarePostFirstProcess}. We notice a clear difference in their distributions. In non-initial processes, 220 out of 394 processes only have one execution wave and do, therefore, not produce any dynamically generated code. In contrast, only 41 samples of 151 in the initial processes have one execution wave, which corresponds to a drop from 56\% in non-initial processes to 27\% in initial processes. The average number of waves in the initial process is $3.39$, where the average number of waves in non-initial processes is $1.64$. 

\subsubsection{Code-size distribution}
The second aspect we consider is how much code the malware executes in each process, and then map this into the entire context of the SPG. To measure this, we collect the number of unique instructions executed in each process and then compare the initial and non-initial processes. The reason we base our measurement on unique instructions and not the total number of executed instructions is to avoid bias because of decryption loops that often have a significantly larger amount of executed instructions in comparison to non-decryption loops. 

\begin{figure}[h]
\centering
\hspace*{-0.8cm} 
\begin{tikzpicture}
\begin{axis}[
    width=0.5\textwidth,
    height=0.4\textwidth,
    xlabel={\% samples},
    ylabel={Instructions executed},
    scaled ticks=false,
    ylabel shift = 1 pt,
    xmin=0, xmax=100,
    ymin=0, ymax=40000,
    legend pos=north west,]
\addplot[
    color=red,]
    coordinates {
(0.000000, 229)  (0.662252, 230)  (1.324503, 230)  (1.986755, 230)  (2.649007, 230)  (3.311258, 230)  (3.973510, 230)  (4.635762, 230)  (5.298013, 230)  (5.960265, 248)  (6.622517, 300)  (7.284768, 300)  (7.947020, 300)  (8.609272, 301)  (9.271523, 303)  (9.933775, 342)  (10.596026, 342)  (11.258278, 342)  (11.920530, 342)  (12.582781, 347)  (13.245033, 406)  (13.907285, 435)  (14.569536, 490)  (15.231788, 623)  (15.894040, 623)  (16.556291, 628)  (17.218543, 655)  (17.880795, 681)  (18.543046, 857)  (19.205298, 890)  (19.867550, 890)  (20.529801, 892)  (21.192053, 892)  (21.854305, 892)  (22.516556, 892)  (23.178808, 892)  (23.841060, 892)  (24.503311, 975)  (25.165563, 1237)  (25.827815, 1239)  (26.490066, 1816)  (27.152318, 1894)  (27.814570, 2027)  (28.476821, 2297)  (29.139073, 2297)  (29.801325, 2297)  (30.463576, 2297)  (31.125828, 2297)  (31.788079, 2297)  (32.450331, 2297)  (33.112583, 2297)  (33.774834, 2297)  (34.437086, 2390)  (35.099338, 2468)  (35.761589, 2490)  (36.423841, 2534)  (37.086093, 2667)  (37.748344, 2697)  (38.410596, 2707)  (39.072848, 2767)  (39.735099, 2776)  (40.397351, 2793)  (41.059603, 2796)  (41.721854, 2879)  (42.384106, 2896)  (43.046358, 2912)  (43.708609, 2963)  (44.370861, 2965)  (45.033113, 3066)  (45.695364, 3165)  (46.357616, 3209)  (47.019868, 3316)  (47.682119, 3319)  (48.344371, 3462)  (49.006623, 3678)  (49.668874, 3683)  (50.331126, 3829)  (50.993377, 3961)  (51.655629, 3961)  (52.317881, 3961)  (52.980132, 3977)  (53.642384, 4088)  (54.304636, 4199)  (54.966887, 4199)  (55.629139, 4209)  (56.291391, 4264)  (56.953642, 4264)  (57.615894, 4267)  (58.278146, 4272)  (58.940397, 4273)  (59.602649, 4353)  (60.264901, 4397)  (60.927152, 4416)  (61.589404, 4440)  (62.251656, 4463)  (62.913907, 4463)  (63.576159, 4463)  (64.238411, 4463)  (64.900662, 4463)  (65.562914, 4487)  (66.225166, 4614)  (66.887417, 4614)  (67.549669, 4734)  (68.211921, 4954)  (68.874172, 4954)  (69.536424, 5032)  (70.198675, 5114)  (70.860927, 5137)  (71.523179, 5138)  (72.185430, 6047)  (72.847682, 6173)  (73.509934, 6175)  (74.172185, 6186)  (74.834437, 6802)  (75.496689, 6964)  (76.158940, 7064)  (76.821192, 7181)  (77.483444, 7189)  (78.145695, 7876)  (78.807947, 8073)  (79.470199, 8082)  (80.132450, 8114)  (80.794702, 8205)  (81.456954, 8811)  (82.119205, 8841)  (82.781457, 9406)  (83.443709, 9724)  (84.105960, 9797)  (84.768212, 10511)  (85.430464, 10633)  (86.092715, 11778)  (86.754967, 11812)  (87.417219, 12184)  (88.079470, 12184)  (88.741722, 12184)  (89.403974, 12371)  (90.066225, 13413)  (90.728477, 13523)  (91.390728, 14113)  (92.052980, 14391)  (92.715232, 15095)  (93.377483, 15684)  (94.039735, 15684)  (94.701987, 15684)  (95.364238, 17898)  (96.026490, 19381)  (96.688742, 20962)  (97.350993, 28758)  (98.013245, 41672)  (98.675497, 71733)  (99.337748, 106454)    };
\addplot[
    color=blue]
    coordinates {
(0.000000, 1)  (0.253807, 1)  (0.507614, 1)  (0.761421, 1)  (1.015228, 15)  (1.269036, 65)  (1.522843, 67)  (1.776650, 67)  (2.030457, 78)  (2.284264, 92)  (2.538071, 112)  (2.791878, 112)  (3.045685, 112)  (3.299492, 112)  (3.553299, 112)  (3.807107, 112)  (4.060914, 112)  (4.314721, 112)  (4.568528, 112)  (4.822335, 112)  (5.076142, 112)  (5.329949, 112)  (5.583756, 112)  (5.837563, 112)  (6.091371, 113)  (6.345178, 113)  (6.598985, 113)  (6.852792, 113)  (7.106599, 113)  (7.360406, 113)  (7.614213, 113)  (7.868020, 113)  (8.121827, 113)  (8.375635, 113)  (8.629442, 113)  (8.883249, 113)  (9.137056, 113)  (9.390863, 113)  (9.644670, 113)  (9.898477, 113)  (10.152284, 113)  (10.406091, 113)  (10.659898, 113)  (10.913706, 113)  (11.167513, 129)  (11.421320, 141)  (11.675127, 141)  (11.928934, 218)  (12.182741, 229)  (12.436548, 229)  (12.690355, 240)  (12.944162, 240)  (13.197970, 240)  (13.451777, 240)  (13.705584, 240)  (13.959391, 252)  (14.213198, 255)  (14.467005, 270)  (14.720812, 277)  (14.974619, 315)  (15.228426, 315)  (15.482234, 315)  (15.736041, 315)  (15.989848, 315)  (16.243655, 315)  (16.497462, 315)  (16.751269, 335)  (17.005076, 335)  (17.258883, 335)  (17.512690, 335)  (17.766497, 335)  (18.020305, 335)  (18.274112, 364)  (18.527919, 364)  (18.781726, 364)  (19.035533, 364)  (19.289340, 364)  (19.543147, 367)  (19.796954, 373)  (20.050761, 374)  (20.304569, 402)  (20.558376, 430)  (20.812183, 430)  (21.065990, 430)  (21.319797, 430)  (21.573604, 430)  (21.827411, 430)  (22.081218, 430)  (22.335025, 430)  (22.588832, 430)  (22.842640, 430)  (23.096447, 430)  (23.350254, 430)  (23.604061, 430)  (23.857868, 430)  (24.111675, 430)  (24.365482, 430)  (24.619289, 430)  (24.873096, 430)  (25.126904, 430)  (25.380711, 430)  (25.634518, 456)  (25.888325, 461)  (26.142132, 461)  (26.395939, 481)  (26.649746, 481)  (26.903553, 481)  (27.157360, 519)  (27.411168, 519)  (27.664975, 519)  (27.918782, 519)  (28.172589, 551)  (28.426396, 551)  (28.680203, 551)  (28.934010, 562)  (29.187817, 583)  (29.441624, 583)  (29.695431, 583)  (29.949239, 583)  (30.203046, 583)  (30.456853, 716)  (30.710660, 729)  (30.964467, 729)  (31.218274, 729)  (31.472081, 731)  (31.725888, 731)  (31.979695, 731)  (32.233503, 731)  (32.487310, 742)  (32.741117, 743)  (32.994924, 764)  (33.248731, 771)  (33.502538, 780)  (33.756345, 810)  (34.010152, 818)  (34.263959, 822)  (34.517766, 822)  (34.771574, 869)  (35.025381, 871)  (35.279188, 878)  (35.532995, 878)  (35.786802, 878)  (36.040609, 878)  (36.294416, 878)  (36.548223, 891)  (36.802030, 891)  (37.055838, 891)  (37.309645, 891)  (37.563452, 891)  (37.817259, 891)  (38.071066, 891)  (38.324873, 891)  (38.578680, 891)  (38.832487, 891)  (39.086294, 891)  (39.340102, 891)  (39.593909, 891)  (39.847716, 891)  (40.101523, 891)  (40.355330, 908)  (40.609137, 908)  (40.862944, 914)  (41.116751, 914)  (41.370558, 923)  (41.624365, 923)  (41.878173, 923)  (42.131980, 962)  (42.385787, 1016)  (42.639594, 1023)  (42.893401, 1308)  (43.147208, 1308)  (43.401015, 1308)  (43.654822, 1309)  (43.908629, 1309)  (44.162437, 1309)  (44.416244, 1309)  (44.670051, 1309)  (44.923858, 1309)  (45.177665, 1309)  (45.431472, 1309)  (45.685279, 1315)  (45.939086, 1334)  (46.192893, 1414)  (46.446701, 1433)  (46.700508, 1433)  (46.954315, 1433)  (47.208122, 1433)  (47.461929, 1462)  (47.715736, 1462)  (47.969543, 1462)  (48.223350, 1503)  (48.477157, 1523)  (48.730964, 1523)  (48.984772, 1543)  (49.238579, 1553)  (49.492386, 1553)  (49.746193, 1553)  (50.000000, 1553)  (50.253807, 1585)  (50.507614, 1585)  (50.761421, 1585)  (51.015228, 1585)  (51.269036, 1585)  (51.522843, 1585)  (51.776650, 1608)  (52.030457, 1608)  (52.284264, 1608)  (52.538071, 1608)  (52.791878, 1608)  (53.045685, 1608)  (53.299492, 1608)  (53.553299, 1608)  (53.807107, 1642)  (54.060914, 1661)  (54.314721, 1661)  (54.568528, 1661)  (54.822335, 1661)  (55.076142, 1680)  (55.329949, 1725)  (55.583756, 1838)  (55.837563, 1862)  (56.091371, 1862)  (56.345178, 1862)  (56.598985, 1862)  (56.852792, 1862)  (57.106599, 1870)  (57.360406, 1870)  (57.614213, 1875)  (57.868020, 1912)  (58.121827, 1912)  (58.375635, 1963)  (58.629442, 1966)  (58.883249, 2008)  (59.137056, 2012)  (59.390863, 2012)  (59.644670, 2059)  (59.898477, 2059)  (60.152284, 2063)  (60.406091, 2063)  (60.659898, 2063)  (60.913706, 2063)  (61.167513, 2093)  (61.421320, 2093)  (61.675127, 2109)  (61.928934, 2116)  (62.182741, 2116)  (62.436548, 2117)  (62.690355, 2126)  (62.944162, 2129)  (63.197970, 2129)  (63.451777, 2137)  (63.705584, 2137)  (63.959391, 2190)  (64.213198, 2261)  (64.467005, 2314)  (64.720812, 2314)  (64.974619, 2314)  (65.228426, 2314)  (65.482234, 2314)  (65.736041, 2430)  (65.989848, 2432)  (66.243655, 2445)  (66.497462, 2452)  (66.751269, 2548)  (67.005076, 2548)  (67.258883, 2693)  (67.512690, 2734)  (67.766497, 2734)  (68.020305, 2773)  (68.274112, 2802)  (68.527919, 2810)  (68.781726, 2853)  (69.035533, 2877)  (69.289340, 2938)  (69.543147, 2938)  (69.796954, 2938)  (70.050761, 3086)  (70.304569, 3086)  (70.558376, 3086)  (70.812183, 3091)  (71.065990, 3144)  (71.319797, 3144)  (71.573604, 3192)  (71.827411, 3244)  (72.081218, 3407)  (72.335025, 3409)  (72.588832, 3410)  (72.842640, 3410)  (73.096447, 3440)  (73.350254, 3496)  (73.604061, 3555)  (73.857868, 3555)  (74.111675, 3561)  (74.365482, 3567)  (74.619289, 3576)  (74.873096, 3582)  (75.126904, 3582)  (75.380711, 3600)  (75.634518, 3608)  (75.888325, 3644)  (76.142132, 3669)  (76.395939, 3669)  (76.649746, 3669)  (76.903553, 3669)  (77.157360, 3680)  (77.411168, 3680)  (77.664975, 3682)  (77.918782, 3715)  (78.172589, 3715)  (78.426396, 3715)  (78.680203, 3715)  (78.934010, 3787)  (79.187817, 3787)  (79.441624, 3792)  (79.695431, 3802)  (79.949239, 3802)  (80.203046, 3962)  (80.456853, 3962)  (80.710660, 3962)  (80.964467, 3962)  (81.218274, 3962)  (81.472081, 3962)  (81.725888, 3984)  (81.979695, 4004)  (82.233503, 4009)  (82.487310, 4089)  (82.741117, 4233)  (82.994924, 4244)  (83.248731, 4379)  (83.502538, 4539)  (83.756345, 4722)  (84.010152, 4722)  (84.263959, 4722)  (84.517766, 4722)  (84.771574, 4729)  (85.025381, 4782)  (85.279188, 4782)  (85.532995, 4782)  (85.786802, 4782)  (86.040609, 4782)  (86.294416, 5230)  (86.548223, 5240)  (86.802030, 5240)  (87.055838, 5362)  (87.309645, 5409)  (87.563452, 5409)  (87.817259, 6121)  (88.071066, 6121)  (88.324873, 6311)  (88.578680, 6312)  (88.832487, 6370)  (89.086294, 6379)  (89.340102, 6625)  (89.593909, 7280)  (89.847716, 7461)  (90.101523, 7633)  (90.355330, 7680)  (90.609137, 7680)  (90.862944, 7787)  (91.116751, 7800)  (91.370558, 7800)  (91.624365, 7809)  (91.878173, 7814)  (92.131980, 7814)  (92.385787, 7814)  (92.639594, 7814)  (92.893401, 7879)  (93.147208, 8126)  (93.401015, 8152)  (93.654822, 8203)  (93.908629, 8301)  (94.162437, 8301)  (94.416244, 8301)  (94.670051, 8301)  (94.923858, 8408)  (95.177665, 8408)  (95.431472, 8408)  (95.685279, 8408)  (95.939086, 8614)  (96.192893, 8745)  (96.446701, 9204)  (96.700508, 9204)  (96.954315, 9243)  (97.208122, 9292)  (97.461929, 9870)  (97.715736, 10437)  (97.969543, 10489)  (98.223350, 11934)  (98.477157, 12249)  (98.730964, 12615)  (98.984772, 12886)  (99.238579, 15864)  (99.492386, 17201)  (99.746193, 17402)  };
\legend{Initial processes, Non-initial processes}
    \end{axis}
\end{tikzpicture}
\caption{The number of instructions executed in initial processes and non-initial processes of all multi-process malware.}
\label{fig:NumberOfInstructionsInMultiProcessMalware}
\end{figure}
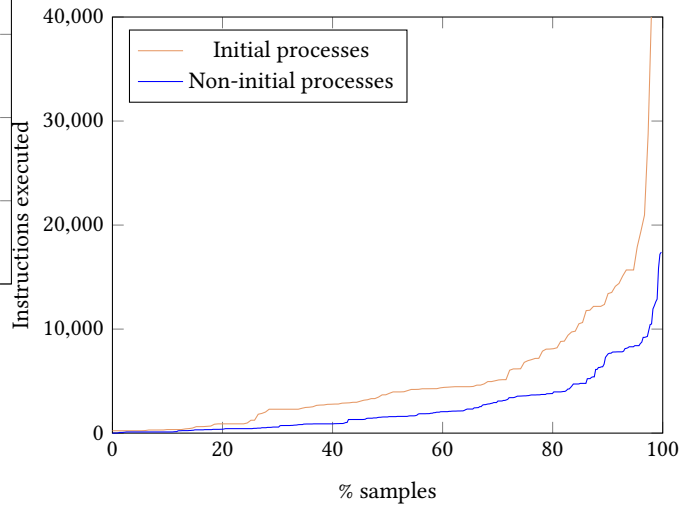

Figure \ref{fig:NumberOfInstructionsInMultiProcessMalware} shows the number of instructions executed in the initial and non-initial processes. The initial processes are significantly more dominant in terms of code size than non-initial processes with an average number of 6157 unique instructions in initial processes over 2500 in non-initial processes. 

\subsubsection{Sensitive API usage}
\begin{figure}[h]
\centering
\begin{tikzpicture}
\begin{axis}[
    ybar,
    enlargelimits=0.15,
    legend style={at={(0.5,-0.15)},
      anchor=north,legend columns=-1},
    ylabel={Average API usage},
    symbolic x coords={Internet,File,Registry, Security},
    xtick=data,
    nodes near coords,
    nodes near coords align={vertical},
    ]
\addplot coordinates {(Internet,0.1) (File,5.0) (Registry,1.5) (Security, 0.1)};
\addplot coordinates {(Internet,0.8) (File,3.6) (Registry,1.4) (Security, 0.9)};

\legend{Initial process,Non-initial process}
\end{axis}
\end{tikzpicture}    
\caption{Sensitive API usage across initial and non-initial processes.}
\label{fig:SensitiveAPIUsage}
\end{figure}
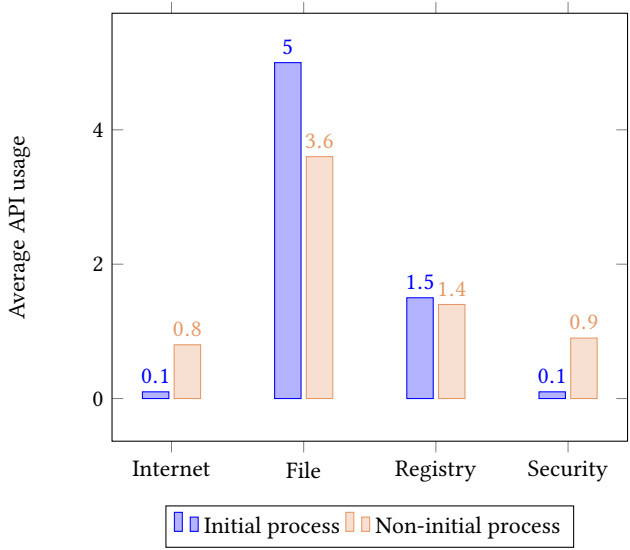

Finally, we look at the distribution of suspicious API calls in samples with multi-process execution. To do this, we generate statistics about where in the SPG malware uses API functions from four groups of sensitive Windows APIs focusing on (1) registry access; (2) internet access; (3) file-system access; and (4) security via process-privilege access. We count the unique number of functions used by the malware from these groups for each process in the SPG and Figure \ref{fig:SensitiveAPIUsage} shows the results for initial and non-initial processes. 

In general, non-initial processes have significantly more usage of APIs related to Internet and security, are on-par with registry-related APIs, but uses less APIs related to the file system. We found 89 non-initial processes call \texttt{ConvertStringSecurityDescriptor\-ToSecurityDescriptor}, which is the most common call from this group, and only six initial processes call this function. We also observe 60 non-initial processes that use \texttt{InitializeSecurityDescri\-ptor}, 60 processes that use \texttt{SetSecurityDescriptorDacl} and 58 processes that use \texttt{GetSecurityDescriptorS\-acl}. All of these functions provide features to change security descriptors of a given process, which is used for process-level privilege elevation. As such, malware is far more likely to initiate privilege escalation techniques in non-initial processes than in the initial process. 

In parallel to security sensitive APIs, we see eight times more usage of Internet-related APIs in non-initial processes than in initial processes. In non-initial processes, we saw a total of 59 processes using the \texttt{WSAStartup} function, which initialises the Windows socket library (\texttt{WS2\_32.dll}) and is the highest level of abstraction for working with Windows sockets. The next most popular Internet-related functions are \texttt{InternetOpen} (27 processes), \texttt{HttpOpenRequest} (22 processes) and \texttt{InternetConnect} (22 processes). In contrast, in initial processes, we only saw 11 use \texttt{WSAStartup} and zero usage of the other functions. 

In Table \ref{tab:FamiliesAndSensitiveAPIs}, we show the number of samples in each family that uses functions from the four sensitive API-groups. The trends from viewing all the samples as a whole follow, and we clearly observe that families most often use Internet and security APIs in non-initial processes. In total, 29 families use Internet-related APIs in non-initial processes whereas only 6 families use them in initial processes, and 18 families use security-related APIs in non-initial processes whereas only seven families use them in initial processes. Roughly half of all samples with multi-process propagation use Internet-related APIs in their non-initial process and a little less than a third uses security-related APIs. This is far more than the usage in initial processes, which is about 7\% for both API groups. We observe slightly more usage of registry-related APIs in non-initial processes and the usage of file system APIs is practically the same between initial and non-initial processes. 

\begin{table}
\scriptsize
\centering  % better than "\begin{center}" and "\end{center}"

\begin{tabular}{|l|c|c|c|c|c|c|c|c|c|}
\hline
        & & \multicolumn{2}{c|}{Internet} & \multicolumn{2}{c|}{Registry} & \multicolumn{2}{c|}{File system} & \multicolumn{2}{c|}{Security} \\
        \hline
        Family & $|\mathcal{M}|$ & Init & Non-init & Init & Non-init  & Init & Non-init & Init & Non-init \\
        \hline
Androm & 2& & 1& 2& 2& 2& 2& & 2\\
Barys & 1& & 1& 1& & 1& 1& & 1\\
Bitman & 1& & & & & 1& 1& & \\
Buzus & 2& & 2& 1& 2& 2& 2& & \\
CTBLocker & 3& & & 3& 3& 3& 3& & 3\\
Cerber & 1& & 1& 1& & 1& 1& 1& \\
CoinMiner & 5& & 2& & 4& 5& 5& 1& \\
Crowti & 2& & 1& 1& & 2& 2& & \\
Cutwail & 2& & & 2& & 2& 2& & \\
Dorkbot & 3& & 2& & 2& 3& 2& & \\
Dridex & 3& & 2& 2& 1& 3& 2& & 1\\
Eldorado & 1& & & & & 1& 1& & \\
Emotet & 9& & 8& 5& 3& 7& 9& & 6\\
Fareit & 1& & 1& & 1& 1& 1& & \\
Fynloski & 2& 1& 1& 2& 1& 2& 1& & \\
Gamarue & 7& & 7& & 7& 7& 7& 1& 7\\
Kasidet & 3& & & 3& 3& 3& 3& & \\
Kovter & 3& & 1& 1& 2& 2& 3& & \\
Kryptik & 3& 1& 2& 1& 1& 3& 3& 1& 1\\
Madangel & 10& & & 2& & 1& 1& & \\
Midie & 4& & 1& 1& 2& 4& 2& & 1\\
Mira & 7& 1& & 1& 7& 7& 7& & \\
Natas & 9& & 5& & 5& 9& 9& & 9\\
Ngrbot & 1& & 1& 1& & 1& & & \\
Nimnul & 6& 1& 3& 2& 3& 6& 6& & 1\\
Nitol & 5& & 3& 5& 1& 5& 3& & 1\\
Nymaim & 2& & 1& 2& 1& 2& 1& & \\
Ramnit & 7& 1& 5& 5& 7& 7& 7& & 1\\
Razy & 9& & 1& 1& 1& 9& 9& & 1\\
Sality & 6& 6& & 6& 1& 6& 6& & \\
Shifu & 1& & 1& & 1& 1& 1& 1& 1\\
Symmi & 3& & 2& 1& 2& 3& 3& & 1\\
TeslaCrypt & 1& & & 1& & 1& 1& & \\
TinyBanker & 9& & 9& 1& 9& 8& 9& & \\
Urausy & 4& & 4& 1& 4& 4& 4& & \\
Ursnif & 3& & & 2& 2& 3& 3& 2& 3\\
VBKrypt & 2& & 2& & 2& & 2& & \\
Vawtrak & 4& & 1& 4& 3& 4& 4& 3& 4\\
Waski & 1& & & 1& 1& 1& 1& & \\
Zbot & 3& & 2& 1& 3& 3& 3& & 2\\
\hline
Total & 151 & 11 & 73 & 63 & 87 & 136 & 133 & 10 & 46 \\
\hline
\end{tabular}    
    \caption{The amount of samples per family that use sensitive functions exposed by the Windows API. The table lists all four categories of sensitive API groups and the amount samples that use these in initial and non-initial processes, respectively. }
    \label{tab:FamiliesAndSensitiveAPIs}
\end{table} 

\subsection{Propagation evolution and inter-family characteristics}
We now move on to the fourth and final research question ``\textit{Has system-wide malware executions changed consistently towards one direction and are there any clear inter-family SPG consistency}''. To answer this question, we first gather statistics about the yearly average of processes involved in malware execution and identify when specific propagation techniques were first used. Following this, we gather statics about inter-family consistency regarding multi-process execution and propagation techniques. 

\subsubsection{Propagation evolution}
\begin{figure}[h]
\centering
\begin{tikzpicture}
\begin{axis}[
    width=0.5\textwidth,
    height=0.4\textwidth,
    ymin=0, ymax=4,
    xtick = {2012, ..., 2018},
    area style,
    xlabel=Year,
    x label style={at={(axis description cs:0.5,-0.2)},anchor=north},
    ylabel=Processes,
    x tick label style={rotate=90, /pgf/number format/1000 sep=}
    ]
\addplot+[ybar, error bars/.cd,
y dir=both,y explicit] plot coordinates {
(2012, 1.805195) +- (0.0, 1.589761/2)
(2013, 1.505747) +- (0.0, 1.310721/2)
(2014, 1.500000) +- (0.0, 1.113885/2)
(2015, 1.756098) +- (0.0, 1.365907/2)
(2016, 1.404255) +- (0.0, 1.247097/2)
(2017, 1.966387) +- (0.0, 2.098962/2)
(2018, 1.293578) +- (0.0, 1.409641/2)};
\end{axis}
\end{tikzpicture}
\caption{Yearly average number of processes and standard deviation of all the samples.}
\label{fig:Average number of processes}
\end{figure}
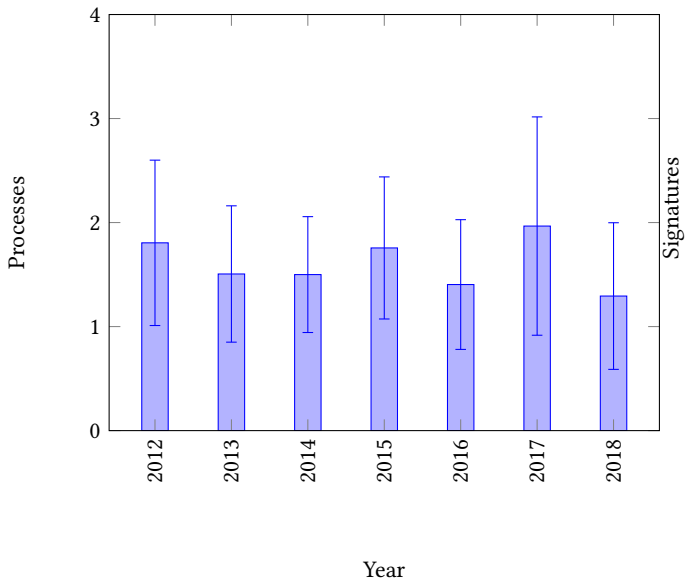

The yearly average and the standard deviation of processes involved in each malware execution are shown in Figure \ref{fig:Average number of processes}. The average number of processes fluctuates, and there is no clear sign towards a steady increase nor decrease. There is roughly an even distribution between the number of processes over the years, and the overall average is 1.61. 

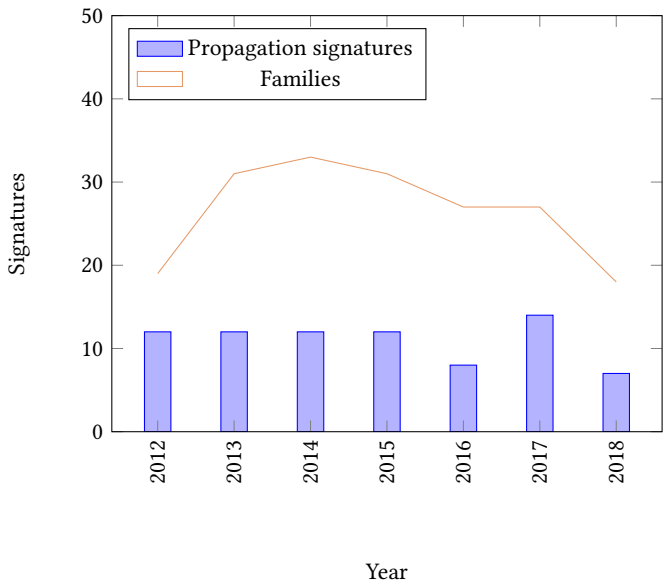
\begin{figure}[h]
\centering
\begin{tikzpicture}
\begin{axis}[
    width=0.5\textwidth,
    height=0.4\textwidth,
    ymin=0, ymax=50,
    xtick = {2012,2013, ..., 2018},
    area style,
    xlabel=Year,
    x label style={at={(axis description cs:0.5,-0.2)},anchor=north},
    ylabel=Signatures,
    x tick label style={rotate=90, /pgf/number format/1000 sep=},
    legend pos=north west
    ]
\addplot+[ybar] plot coordinates {
(2012,12)
(2013,12)
(2014,12)
(2015,12)
(2016,8)
(2017,14)
(2018,7)
};

\addplot[
    color=red]
    coordinates {
(2012, 19)
(2013, 31)
(2014, 33)
(2015, 31)
(2016, 27)
(2017, 27)
(2018, 18)};
\legend{Propagation signatures, Families}
\end{axis}
\end{tikzpicture}
\caption{Number of different propagation signatures each year and the number of families each year.}
\label{fig:NumberOfSignaturesEachYear}
\end{figure}

\begin{figure}[h]
\centering
\hspace*{-1.0cm} 
\begin{tikzpicture}
\begin{axis}[
    width=0.5\textwidth,
    height=0.4\textwidth,
    ymin=0, ymax=13,
    xtick = {2012,2013, ..., 2018},
    area style,
    xlabel=Year,
    x label style={at={(axis description cs:0.5,-0.2)},anchor=north},
    ylabel=Signatures,
    x tick label style={rotate=90, /pgf/number format/1000 sep=},
    legend pos=north west
    ]
\addplot+[ybar] plot coordinates {
(2012, 12)
(2013, 7)
(2014, 6)
(2015, 3)
(2016, 1)
(2017, 2)
(2018, 2)};
\end{axis}
\end{tikzpicture}
\caption{The years signatures were first used.}
\label{fig:SignatureDiscoveryByYear}
\end{figure}
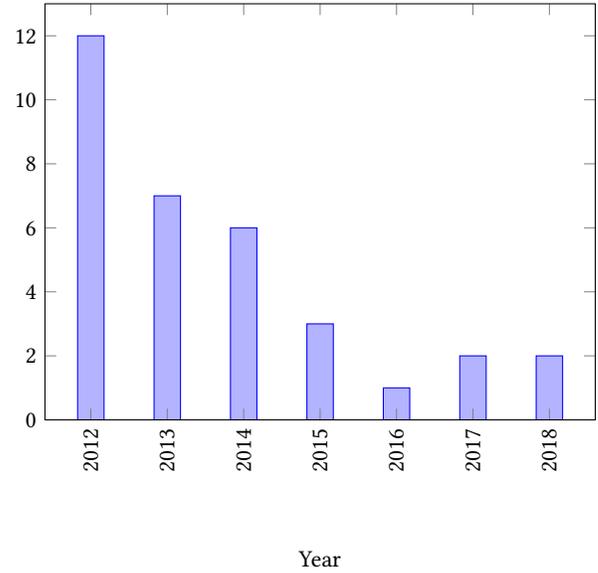

Next, we investigate how the multi-process propagation techniques have evolved by correlating the specific signatures we developed to recognise multi-process propagation in Section \ref{sec:signature_diversity} to when they were first used. The number of unique signatures found each year and the number of families in our data set for each year are shown in Figure \ref{fig:NumberOfSignaturesEachYear}. In general, we observe a steady usage of signatures ranging from seven at the lowest (2018) to fourteen at the highest (2017). In Figure \ref{fig:SignatureDiscoveryByYear}, we show the number of signatures that were discovered a given year, and, interestingly, we observe that the number of novel propagation techniques is steadily decreasing. This decrease shows clearly in that 25 out of 33 injection signatures were first used in 2012, 2013 and 2014, which amounts to 75\% of all injections being invented before 2015. 

\subsubsection{Inter-family propagation characteristics}
\label{chapter6:InterFamilyConsistencies}
Throughout this paper, we have explored various family-related aspects of system-wide malware execution, and we now continue in this domain by giving a more general assessment of inter-family consistency using several relevant metrics. We continue with our definitions from earlier and denote the number of samples in each family as $|\mathcal{S}|$ and the number of samples that perform multi-process propagation as $|\mathcal{M}|$. 

The first consistency measurement we determine is whether families are stable in terms of multi-process propagation. To do this, we compute the maximum number of samples in each family that either perform multi-process propagation or stays in the same process over the total amount of samples, given as $\frac{max(|\mathcal{M}|, |\mathcal{S}| - |\mathcal{M}|)}{|\mathcal{S}|}$. This is shown in the fourth column from the left in Table \ref{tab:malware_family_intrinsics}, and we find an overall consistency average of $0.86$. We do the same check but restricted to the data set of families with samples that do multi-process propagation, i.e. $\frac{|\mathcal{M}|}{|\mathcal{S}|}$ for families with $|\mathcal{M}| \neq 0$, and the result is shown in the fifth left-most column in Table \ref{tab:malware_family_intrinsics}. The overall average decreases by $0.48$ to $0.38$. This is unexpected since it shows there is little consistency in the number of samples that perform system-wide propagation within malware families where at least one sample performs multi-process propagation.

Next, we consider the number of unique signatures deployed in each malware family, which we denote $\Sigma$. We observe an average of $2.7$ different signatures in malware families that have multi-process propagating samples, as shown in the sixth column from the left in Table \ref{tab:malware_family_intrinsics}. Another interesting measurement is the consistency only between the malware samples that perform multi-process propagation. To quantify this, we count the number of samples in each family that has the same set of propagation signatures, and we call this $\Sigma_{eq}$. For families with multi-process propagation, the average number of samples with the same set of signatures is $2.62$. To measure the inter-family consistency we then divide $\Sigma_{eq}$ with the number of propagating samples from each family, given as $\frac{\Sigma_{eq}}{|\mathcal{M}|}$ and this is shown in the second column from the right. Here, we see an overall average of $0.73$, meaning 73\% of samples in families with propagating samples have the same set of signatures.

Finally, we look at consistency based on the maximum number of samples with similar signatures and the number of samples without multi-process propagation, given as $\mathcal{S}_{eq} = max(\Sigma_{eq}, |\mathcal{S}| - |\mathcal{M}|)$. From $\mathcal{S}_{eq}$ we then calculate $\frac{\mathcal{S}_{eq}}{|\mathcal{S}|}$ to get a consistency measurement on the entire family based on samples with the same set of signatures or no signatures at all. On average, we see a $0.83$ consistency as shown in the rightmost column of Table \ref{tab:malware_family_intrinsics}. This shows that 83\% of samples in each family either has the same set of signatures for multi-process propagation, or no samples in the family have multi-process propagation at all. We found no families that have multi-process propagation to have a perfect score in this context. 

\begin{table*}{}
\centering
\small
    \begin{tabular}{l|c|c|c|c|c||c|c||c|c}
    \toprule
    \hline
         Family & $|\mathcal{S}|$ & $|\mathcal{M}|$ & $\frac{max(|\mathcal{M}|, |\mathcal{S}| - |\mathcal{M}|)}{|\mathcal{S}|}$ & $\frac{|\mathcal{M}|}{|\mathcal{S}|}$ & $\Sigma$ & $\Sigma_{eq}$ & $\mathcal{S}_{eq}$  & $\frac{\Sigma_{eq}}{|\mathcal{M}|}$ &  $\frac{\mathcal{S}_{eq}}{|\mathcal{S}|}$\\
    \hline
Androm &  	10 & 	2 & 	8.00 & 	0.20 &  	3 &  1 & 8 & 0.50 &  0.80 \\
Artemis &  	10 & 	0 & 	10.00 & 	--- & 	--- &  --- &  10 & --- &  1.00 \\
Barys &  	10 & 	1 & 	9.00 & 	0.10 &  	1 &  1 & 9 & 1.00 &  0.90 \\
Bitman &  	10 & 	1 & 	9.00 & 	0.10 &  	1 &  1 & 9 & 1.00 &  0.90 \\
Buzus &  	10 & 	2 & 	8.00 & 	0.20 &  	7 &  1 & 8 & 0.50 &  0.80 \\
CTBLocker &  	10 & 	3 & 	7.00 & 	0.30 &  	1 &  3 & 7 & 1.00 &  0.70 \\
Cerber &  	10 & 	1 & 	9.00 & 	0.10 &  	2 &  1 & 9 & 1.00 &  0.90 \\
CoinMiner &  	10 & 	5 & 	5.00 & 	0.50 &  	5 &  2 & 5 & 0.40 &  0.50 \\
CosmicDuke &  	10 & 	0 & 	10.00 & 	--- & 	--- &  --- &  10 & --- &  1.00 \\
Crowti &  	10 & 	2 & 	8.00 & 	0.20 &  	1 &  2 & 8 & 1.00 &  0.80 \\
Cryptlock &  	10 & 	0 & 	10.00 & 	--- & 	--- &  --- &  10 & --- &  1.00 \\
Cutwail &  	10 & 	2 & 	8.00 & 	0.20 &  	1 &  2 & 8 & 1.00 &  0.80 \\
DealPly &  	10 & 	0 & 	10.00 & 	--- & 	--- &  --- &  10 & --- &  1.00 \\
Dorkbot &  	10 & 	3 & 	7.00 & 	0.30 &  	3 &  1 & 7 & 0.33 &  0.70 \\
Dridex &  	10 & 	3 & 	7.00 & 	0.30 &  	3 &  1 & 7 & 0.33 &  0.70 \\
Eldorado &  	10 & 	1 & 	9.00 & 	0.10 &  	1 &  1 & 9 & 1.00 &  0.90 \\
Emotet &  	10 & 	9 & 	9.00 & 	0.90 &  	6 &  5 & 5 & 0.56 &  0.50 \\
Fareit &  	10 & 	1 & 	9.00 & 	0.10 &  	1 &  1 & 9 & 1.00 &  0.90 \\
Flood &  	10 & 	0 & 	10.00 & 	--- & 	--- &  --- &  10 & --- &  1.00 \\
Fujacks &  	10 & 	0 & 	10.00 & 	--- & 	--- &  --- &  10 & --- &  1.00 \\
Fynloski &  	10 & 	2 & 	8.00 & 	0.20 &  	2 &  1 & 8 & 0.50 &  0.80 \\
Gamarue &  	10 & 	7 & 	7.00 & 	0.70 &  	1 &  7 & 7 & 1.00 &  0.70 \\
Gootkit &  	10 & 	0 & 	10.00 & 	--- & 	--- &  --- &  10 & --- &  1.00 \\
Kasidet &  	10 & 	3 & 	7.00 & 	0.30 &  	1 &  3 & 7 & 1.00 &  0.70 \\
Kazy &  	10 & 	0 & 	10.00 & 	--- & 	--- &  --- &  10 & --- &  1.00 \\
Kovter &  	10 & 	3 & 	7.00 & 	0.30 &  	3 &  1 & 7 & 0.33 &  0.70 \\
Kraddare &  	10 & 	0 & 	10.00 & 	--- & 	--- &  --- &  10 & --- &  1.00 \\
Kryptik &  	10 & 	3 & 	7.00 & 	0.30 &  	3 &  1 & 7 & 0.33 &  0.70 \\
Madangel &  	10 & 	10 & 	10.00 & 	1.00 &  	2 &  8 & 8 & 0.80 &  0.80 \\
Madi &  	10 & 	0 & 	10.00 & 	--- & 	--- &  --- &  10 & --- &  1.00 \\
Mamba &  	10 & 	0 & 	10.00 & 	--- & 	--- &  --- &  10 & --- &  1.00 \\
Mazam &  	10 & 	0 & 	10.00 & 	--- & 	--- &  --- &  10 & --- &  1.00 \\
Midie &  	10 & 	4 & 	6.00 & 	0.40 &  	2 &  3 & 6 & 0.75 &  0.60 \\
MiniDuke &  	10 & 	0 & 	10.00 & 	--- & 	--- &  --- &  10 & --- &  1.00 \\
Mira &  	10 & 	7 & 	7.00 & 	0.70 &  	1 &  7 & 7 & 1.00 &  0.70 \\
Natas &  	10 & 	9 & 	9.00 & 	0.90 &  	4 &  4 & 4 & 0.44 &  0.40 \\
Neshta &  	10 & 	0 & 	10.00 & 	--- & 	--- &  --- &  10 & --- &  1.00 \\
Neshuta &  	10 & 	0 & 	10.00 & 	--- & 	--- &  --- &  10 & --- &  1.00 \\
Ngrbot &  	10 & 	1 & 	9.00 & 	0.10 &  	1 &  1 & 9 & 1.00 &  0.90 \\
Nimnul &  	10 & 	6 & 	6.00 & 	0.60 &  	6 &  3 & 4 & 0.50 &  0.40 \\
Nitol &  	10 & 	5 & 	5.00 & 	0.50 &  	1 &  5 & 5 & 1.00 &  0.50 \\
Nymaim &  	10 & 	2 & 	8.00 & 	0.20 &  	1 &  2 & 8 & 1.00 &  0.80 \\
Otwycal &  	10 & 	0 & 	10.00 & 	--- & 	--- &  --- &  10 & --- &  1.00 \\
Padodor &  	10 & 	0 & 	10.00 & 	--- & 	--- &  --- &  10 & --- &  1.00 \\
Parite &  	10 & 	0 & 	10.00 & 	--- & 	--- &  --- &  10 & --- &  1.00 \\
Pony &  	10 & 	0 & 	10.00 & 	--- & 	--- &  --- &  10 & --- &  1.00 \\
Pronny &  	10 & 	0 & 	10.00 & 	--- & 	--- &  --- &  10 & --- &  1.00 \\
Ramnit &  	10 & 	7 & 	7.00 & 	0.70 &  	7 &  3 & 3 & 0.43 &  0.30 \\
Razy &  	10 & 	9 & 	9.00 & 	0.90 &  	3 &  7 & 7 & 0.78 &  0.70 \\
Renos &  	10 & 	0 & 	10.00 & 	--- & 	--- &  --- &  10 & --- &  1.00 \\
Rovnix &  	10 & 	0 & 	10.00 & 	--- & 	--- &  --- &  10 & --- &  1.00 \\
Sality &  	10 & 	6 & 	6.00 & 	0.60 &  	1 &  6 & 6 & 1.00 &  0.60 \\
Shifu &  	10 & 	1 & 	9.00 & 	0.10 &  	1 &  1 & 9 & 1.00 &  0.90 \\
Simda &  	10 & 	0 & 	10.00 & 	--- & 	--- &  --- &  10 & --- &  1.00 \\
Symmi &  	10 & 	3 & 	7.00 & 	0.30 &  	4 &  1 & 7 & 0.33 &  0.70 \\
TeslaCrypt &  	10 & 	1 & 	9.00 & 	0.10 &  	1 &  1 & 9 & 1.00 &  0.90 \\
TinyBanker &  	10 & 	9 & 	9.00 & 	0.90 &  	3 &  8 & 8 & 0.89 &  0.80 \\
Urausy &  	10 & 	4 & 	6.00 & 	0.40 &  	5 &  2 & 6 & 0.50 &  0.60 \\
Ursnif &  	10 & 	3 & 	7.00 & 	0.30 &  	6 &  1 & 7 & 0.33 &  0.70 \\
VBKrypt &  	10 & 	2 & 	8.00 & 	0.20 &  	3 &  2 & 8 & 1.00 &  0.80 \\
Vawtrak &  	10 & 	4 & 	6.00 & 	0.40 &  	3 &  2 & 6 & 0.50 &  0.60 \\
Wannacry &  	10 & 	0 & 	10.00 & 	--- & 	--- &  --- &  10 & --- &  1.00 \\
Waski &  	10 & 	1 & 	9.00 & 	0.10 &  	1 &  1 & 9 & 1.00 &  0.90 \\
Zbot &  	10 & 	3 & 	7.00 & 	0.30 &  	6 &  1 & 7 & 0.33 &  0.70 \\
vilsel &  	10 & 	0 & 	10.00 & 	--- & 	--- &  --- &  10 & --- &  1.00 \\
\hline
Total average & 10.00 & 2.32 & 0.86 & 0.38 & 2.70 & 2.62 & 8.26 & 0.73 & 0.83\\
\bottomrule
\hline
    \end{tabular}
    \caption{Inter-family malware propagation characteristics.}
    \label{tab:malware_family_intrinsics}
\end{table*}

\section{Discussion}
\label{chapter6_discussion_of_results}
We now give answers to our research questions, discuss the results in more depth and also discuss the limitations of our study. We present our discussion in chronological order of the research questions.  \\

\subsection{Answers to research questions}
\textbf{Research question 1.}\textit{ Is system-wide malware propagation prevalent? Yes!}\\
We find that 23.23\% of samples perform multi-process propagation, showing that almost a quarter of malware samples rely on host-based propagation. In contrast, PaloAlto reports that 13.5\% \cite{PaloAltoReport} of malware samples in 2013 contain code injections and Ugarte et al. \cite{Ugarte-pedrero_sok:deep} report that 15.6\% samples (1,213 of 7,729) contain multi-process execution in a data set spanning mid-2007 to mid-2014. We believe the main reason for this is that our system is more general and captures multi-process execution that previous work miss. Another possibility for different results is data set difference, in that Ugarte et al. rely on samples that are older than ours. 

System-wide malware propagation is not only prevalent in terms of multi-process propagation but also the number of dynamically generated execution waves. We find that 60\% of all samples have multiple execution waves, meaning they deploy some form of dynamically generated code. In contrast, Ugarte et al. \cite{Ugarte-pedrero_sok:deep} found 78.7\% (6088 out of 7729 samples), Codisasm \cite{Bonfante:2015:CMS:2810103.2813627} reports 68\% and Kang et al. \cite{Kang:2007:RHC:1314389.1314399} report 98\% (367 out of 374). Interestingly, we report the lowest number since our model is generic in terms of system-wide malware execution. However, our model also distinguishes malicious and benign code execution more precisely, and we suspect this to be the main reason. Previous work unreasonably over-approximates some aspects of the execution, e.g. capture dynamically generated code by any code in a process with malware code executing and these over-approximations may capture non-malicious execution waves where we are highly resistant to this. 

System-wide malware propagation has also been prevalent for many years, showing that indeed this is not a new issue. We observe a steady use of multi-process propagation throughout our entire data set, spanning 2012-2018, and we find that 40 out of a total 65 families use multi-process propagation, which is roughly two-thirds of the families in our data set. As such, system-wide propagation is prevalent not only in the number of total samples but also in the majority of families. \\

\textbf{Research question 2.} \textit{Are the propagation strategies used in the wild diverse? Yes!}\\
We find that there is a range of 2 to 11 processes in which malware with system-wide propagation execute. As such, there is a significant difference directly in the number of processes, although 85\% of samples with multi-process execution execute in either two or three processes. In addition to this, there is also a broad range of execution waves deployed by malware samples, with the majority ranging between two and 25 execution waves. We find only 13 samples with more than 25 execution waves. Previous work also reports a diversity in the number of process executions and the number of execution waves \cite{Bonfante:2015:CMS:2810103.2813627, Dinaburg:2008:EMA:1455770.1455779, Kang:2007:RHC:1314389.1314399, Ugarte-pedrero_sok:deep}. 

Furthermore, we also observe diversity in process-depth, wave-depth and SPG-width. We see a maximum process-depth of five processes and a maximum SPG-width of seven processes. This means we see several malware samples that both spread out across several processes and also perform their system-wide propagation in a chain-like fashion. 

One of the areas that show significant diversity in propagation strategy is the specific techniques malware use to propagate across processes. We find a total of 33 different propagation techniques, and most of these have less than ten samples using the given technique. However, many of the more rarely-seen techniques are derivatives of popular techniques. These are derivatives in various ways such as using hooking to rely on only a subset of the standard APIs. In addition to this, there is also an even distribution between propagations that inject code into existing processes and propagations that create new processes.

Interestingly, we also find that malware with samples that do system-wide propagation use on average $2.6$ different propagation techniques. This also means there is diversity within the SPG of each malware sample, in the specific way they propagate through the system. 

We also observe that initial processes execute more code than non-initial processes. On average, the initial processes execute about three times more unique instructions than non-initial processes. Furthermore, the set of processes that malware target to perform multi-process propagation is mainly divided between a small number of standard Windows applications.  \\

\textbf{Research question 3.} \textit{Are there clear relations between malicious behaviour and system-wide malware propagation? Yes!}\\
Two key observations provide arguments for the answer to our third research question. First, the number of execution waves in the initial process significantly outnumbers the number of execution waves in non-initial processes. In total, 73\% of samples that deploy multi-process propagation use dynamically generated code in their initial process where the figure is 44\% for processes that are a result of multi-process propagation. As such, dynamically generated code is a technique that is mostly used in the initial process but occurs in almost half of the non-initial processes.  

Second, the use of sensitive API calls is a central feature that distinguishes code in the initial process and code in non-initial processes. We find far more usage of security-related APIs and Internet-related APIs in non-initial processes. This makes sense because malware uses propagation techniques, including dynamically generated code and multi-process propagation, to hide the malicious code and make the analysis of it more complicated. In addition to this, it makes sense that sensitive API calls are mainly called in non-initial processes because a large number of targets are benign Windows programs where sensitive API calls are considered less suspicious. \\

\textbf{Research question 4.}\textit{ Has system-wide malware executions changed consistently towards one direction and are there any clear inter-family SPG consistency?} \textit{We find no evidence of consistent changes over the years, and for families, we find areas of both inter-family consistency and diversity!}\\
We find no clear evidence that the use of system-wide propagation in malware is increasing nor decreasing but rather observe a fluctuating trend. Specifically, there is no increase in either the number of processes used by malware or execution waves and there is no increase in the number of novel API patterns used to propagate to multiple processes. We find that there is a constant diversity amongst signatures used every year, but that the number of unique signatures invented is steadily decreasing. 

From a high-level perspective, we find several metrics that show inter-family consistency. We find that 86\% of samples in each family either propagate or do not propagate and, closely related, we find that 83\% of samples in each family follow one direction of either not deploying any system-wide propagation or implementing the same set of propagation approaches. Additionally,  In 17 out of 40 families with multi-process execution, we find that the targets of the multi-process samples of each family are the same, albeit it is only six of these families that have multiple multi-process samples. In terms of propagation strategy, we find that for families with multi-process execution an average of 73\% of the samples use the same APIs to propagate through the system, and we find several families where the samples in the respective family propagate using a single signature. Furthermore, we find that in 27 out of 31 families that use Internet-related APIs the samples of each family purely use functions from the API via initial or non-initial processes, meaning the samples in each family consistently use Internet-related APIs in the same way. We find a somewhat similar trend with APIs related to process-level privileges where samples from 12 families, out of a total 17, exclusively use the API from either initial or non-initial processes, albeit, 8 of these only have a single sample using the security-related APIs. 

However, we also observe several metrics that show signs of inter-family diversity. For example, in the 40 families that have samples with multi-process execution, we find an inter-family average of 38\% samples exhibiting multi-process execution. Furthermore, it is only in one family where all samples deploy multi-process propagation and only in 10 families that more than half of the samples deploy multi-process execution. Additionally, we find a significant variation in 9 out of all 65 families in terms of the number of execution waves deployed by the samples of the respective families, in that the standard deviation is larger than the average number of execution waves. Similarly, we find that 17 out of 40 families with multi-process execution have varying levels of process-depths, meaning the samples in each family deploy diverse propagation strategies.

\subsection{Limitations}
The main limitation of our study is that we execute the samples under one specific setting. To get broader insights about malware propagation we can leverage the use of differential studies, by analysing the malware under different contexts, similar to Cozzi et al. \cite{UnderstandingLinuxMalware}. For example, we set the recording time to 25 seconds, and we did not perform any user stimulation. Increasing the recording time is likely to produce interesting results, and supplying user stimulation can trigger more behaviours in some types of malware. Another promising avenue is changing parameters in the execution environment, such as the state of the guest machine. For example, we used a vanilla Windows 7 with no processes running except for the standard Windows applications. However, some malware samples rely on injecting into specific processes, such as web browsers, and the state of our guest machine does not enable this behaviour. By providing a more realistic setup, we may gather more resourceful results. In addition to this, we could also collect more comprehensive statistics by extending Minerva to support 64-bit architectures, as some malware samples are observed to deploy exotic propagation techniques only on this architecture \cite{DridexV4Atombinging}.

\section{Related work}
\textbf{Large-scale dynamic analysis studies.} There are several existing studies based on large-scale dynamic analysis of malware. Bayer et al. performed a study \cite{DBLP:conf/leet/BayerHBK09} on samples from early 2007 to late 2008 based on their Anubis malware analysis platform. Their data set is based on submissions to their web portal, and they investigate various aspects such as file system, network and registry activity as well as botnet activity and sandbox detection. Ugarte et al. \cite{Ugarte-pedrero_sok:deep} presents a large-scale study on packers based on an analysis platform build on top of TEMU. However, their system is not as general as Minerva in terms of capturing system-wide malware execution \cite{2019arXiv190809204K} and their works is largely focused on creating a packer taxonomy and analysing their data set in relation to this taxonomy. Severi at al. presents the Malrec malware analysis system \cite{DBLP:conf/dimva/SeveriLD18}, which is also based on the PANDA instrumentation environment, and their study incorporates 66,301 samples spread over 1,270 families as identified by AVClass \cite{10.1007/978-3-319-45719-2_11}. They first perform a study on how much malware modify kernel code as a means of detecting privilege escalation and then performed a study on malware classification.  \\

\textbf{Other large-scale malware analysis studies.} Plohmann et al. present the Malpedia platform \cite{CybIN}, which is a collaborative effort to gather samples and structure the malware landscape. They collect a corpus of 1800 malware samples spanning 600 families and have manually unpacked many of them. Based on the data set, they perform a study on various elements collected through static analysis such as the PE headers, control-flow analysis and API usage. Although the majority of large-scale studies focus on Windows malware, there is some work about other operating systems and architectures. A recent study by Cozzi et al. \cite{EURECOM+5489} performs a comprehensive investigation into Linux malware. The authors collect a data set of 10,548 Linux malware samples for more than eight different architectures and study several aspects like ELF header manipulation, persistence, deception, privilege escalation and process interactions. The growing number of Android and mobile phones have also motivated several large-scale studies for malware on the Android platform \cite{Felt:2011:SMM:2046614.2046618, DBLP:journals/csur/TamFASC17, 10.1007/978-3-319-60876-1_12, Zhou:2012:DAM:2310656.2310710}. 

\section{Conclusion}
In this paper, we performed a large-scale analysis of malware in the wild, spanning 650 samples and 65 different families. Our study focused on the three aspects of system-wide malware propagation: (1) prevalence and diversity; (2) relationship to malicious behaviours; and (3) evolution and inter-family consistency. 

We collected vast amounts of statistics to derive insights about malware propagation and discussed our results in detail. We found that system-wide propagation is prevalent and diverse amongst malware samples. We found clear relationships between propagation and malware behaviours and found mixed signals in terms of inter-family consistency. Surprisingly we did not see any increase in malware propagation through the years but instead observed a steady and consistent use. 

The results of our study show that we can use a carefully selected data set to derive a high-level view of malware propagation, albeit this view can be blurry in places. We used this high-level view to identify key characteristics of malware propagation that have implications for existing work and also opens opportunities for new research avenues.

%%
%% The next two lines define the bibliography style to be used, and
%% the bibliography file.
\bibliographystyle{ACM-Reference-Format}
\bibliography{sample-base}

%%
%% If your work has an appendix, this is the place to put it.
\appendix

\section{Data tables}

\begin{table*}
\small
\centering  % better than "\begin{center}" and "\end{center}"
\begin{tabular}{|l|c|P{1.080}|c|c|c|}
\hline
        Family & $|\mathcal{M}|$ & (target process name, sample count), ...,  & $\mathcal{I}$ & $\frac{|\mathcal{I}|}{|\mathcal{M}|}$ & $\frac{|\mathcal{I}'|}{|\mathcal{T}|}$ \\
\hline Androm & 2 &(sample.exe,2), (iexplore.exe,1), (uhUhRBmr.exe,1) & 1 & 0.50 & 0.50\\
\hline Barys & 1 &(explorer.exe,1), (svchost.exe,1) & 1 & \textbf{1.00} & \textbf{1.00}\\
\hline Bitman & 1 &(sample.exe,1) & 1 & \textbf{1.00} & \textbf{1.00}\\
\hline Buzus & 2 &(explorer.exe,2), (sample.exe,2), (iexplore.exe,2) & 2 &\textbf{ 1.00} & \textbf{1.00}\\
\hline CTBLocker & 3 &(sample.exe,3), (iexplore.exe,3) & 3 & \textbf{1.00 }&\textbf{ 1.00}\\
\hline Cerber & 1 &(ns710D.tmp,1), (timetasks.exe,1) & 1 & \textbf{1.00} & \textbf{1.00}\\
\hline CoinMiner & 5 &(sample.tmp,2), (ns9204.tmp,1), (CNminer.exe,1), (DbQZbGtq.exe,1), (WMIC.exe,1), (ns97DF.tmp,1) & 1 & 0.20 & 0.43\\
\hline Crowti & 2 &(explorer.exe,2), (svchost.exe,1) & 1 & 0.50 & 0.67\\
\hline Cutwail & 2 &(sample.exe,2), (iexplore.exe,2) & 2 & \textbf{1.00} &\textbf{ 1.00}\\
\hline Dorkbot & 3 &(sample.exe,3), (taskhost.exe,2), (conhost.exe,2), (explorer.exe,2), (dwm.exe,2), (cmd.exe,2) & 2 & 0.67 & 0.92\\
\hline Dridex & 3 &(explorer.exe,2), (svchost.exe,1), (edg7AF9.exe,1) & 1 & 0.33 & 0.50\\
\hline Eldorado & 1 &(708D.tmp,1) & 1 & \textbf{1.00} & \textbf{1.00}\\
\hline Emotet & 9 &(sample.exe,8), (explorer.exe,8), (svchost.exe,5), (iexplore.exe,4), (taskhost.exe,3), (conhost.exe,3), (dwm.exe,3), (cmd.exe,3), (winver.exe,2) & 4 & 0.44 & 0.41\\
\hline Fareit & 1 &(sample.exe,1) & 1 & \textbf{1.00} & \textbf{1.00}\\
\hline Fynloski & 2 &(notepad.exe,1), (sample.exe,1) & 1 & 0.50 & 0.50\\
\hline Gamarue & 7 &(wuauclt.exe,7) & 7 & \textbf{1.00} & \textbf{1.00}\\
\hline Kasidet & 3 &(explorer.exe,3) & 3 & \textbf{1.00} & \textbf{1.00}\\
\hline Kovter & 3 &(sample.exe,2), (svchost.exe,2) & 2 & 0.67 & 0.50\\
\hline Kryptik & 3 &(explorer.exe,1), (dllhost.exe,1), (73C8.tmp,1) & 1 & 0.33 & 0.33\\
\hline Madangel & 10 &(sample.exe,10), (explorer.exe,2) & 10 & \textbf{1.00} & 0.83\\
\hline Midie & 4 &(sample.exe,2), (sample.usr,2), (iexplore.exe,1) & 2 & 0.50 & 0.40\\
\hline Mira & 7 &(xvkwym.exe,3), (raphw.exe,2), (vlfsay.exe,2) & 3 & 0.43 & 0.43\\
\hline Natas & 9 &(sample.exe,9), (taskhost.exe,5), (conhost.exe,5), (explorer.exe,5), (cmd.exe,5), (dwm.exe,5), (ahos.exe,1), (uluwo.exe,1), (zaar.exe,1), (yros.exe,1), (loid.exe,1) & 5 & 0.56 & 0.77\\
\hline Ngrbot & 1 &(explorer.exe,1) & 1 & \textbf{1.00} & \textbf{1.00}\\
\hline Nimnul & 6 &(iexplore.exe,2), (DesktopLayer.e,2), (sampleSrv.exe,2), (tXoPUA.exe,1), (sample.exe,1), (OZpdVg.exe,1), (uhUhRBmr.exe,1), (fuDwxVB.exe,1) & 2 & 0.33 & 0.55\\
\hline Nitol & 5 &(sample.exe,5), (iexplore.exe,5) & 5 & \textbf{1.00} & \textbf{1.00}\\
\hline Nymaim & 2 &(sample.exe,2), (iexplore.exe,2) & 2 & \textbf{1.00} & \textbf{1.00}\\
\hline Ramnit & 7 &(iexplore.exe,6), (explorer.exe,5), (samplemgr.exe,3), (DesktopLayer.e,3), (sampleSrv.exe,2), (conhost.exe,1), (OXYbHl.exe,1), (rundll32.exe,1), (OXYbHlSrv.exe,1), (taskhost.exe,1), (dwm.exe,1), (cmd.exe,1) & 3 & 0.43 & 0.35\\
\hline Razy & 9 &(750F.tmp,1), (sample.exe,1), (79D0.tmp,1), (72FD.tmp,1), (iexplore.exe,1), (7686.tmp,1), (7425.tmp,1), (7722.tmp,1), (7119.tmp,1), (73E7.tmp,1) & 1 & 0.11 & 0.20\\
\hline Sality & 6 &(taskhost.exe,6), (conhost.exe,6), (explorer.exe,6), (dwm.exe,6), (cmd.exe,6), (rundll32.exe,4), (sample.tmp,1) & 6 & \textbf{1.00} & 0.86\\
\hline Shifu & 1 &(taskhost.exe,1), (conhost.exe,1), (explorer.exe,1), (dwm.exe,1), (cmd.exe,1), (dllhost.exe,1) & 1 & \textbf{1.00} & \textbf{1.00}\\
\hline Symmi & 3 &(sample.exe,3), (explorer.exe,1), (svchost.exe,1), (uhUhRBmr.exe,1) & 1 & 0.33 & 0.50\\
\hline TeslaCrypt & 1 &(sample.exe,1) & 1 & \textbf{1.00} & \textbf{1.00}\\
\hline TinyBanker & 9 &(winver.exe,9), (explorer.exe,9), (taskhost.exe,2), (conhost.exe,2), (dwm.exe,2), (cmd.exe,2), (sample.exe,1) & 9 & \textbf{1.00} & 0.67\\
\hline Urausy & 4 &(sample.exe,4), (explorer.exe,3), (svchost.exe,3) & 3 & 0.75 & 0.90\\
\hline Ursnif & 3 &(explorer.exe,3), (cliciles.exe,1), (chtborui.exe,1), (~7F9A.tmp,1), (factura.exe,1) & 1 & 0.33 & 0.43\\
\hline VBKrypt & 2 &(sample.exe,2), (taskhost.exe,2), (conhost.exe,2), (winver.exe,2), (explorer.exe,2), (dwm.exe,2), (cmd.exe,2) & 2 & \textbf{1.00} & \textbf{1.00}\\
\hline Vawtrak & 4 &(explorer.exe,4), (taskhost.exe,3), (conhost.exe,3), (dwm.exe,3), (cmd.exe,3), (mainOUT-crypt,1), (svchost.exe,1) & 3 & 0.75 & 0.83\\
\hline Waski & 1 &(sample.tmp,1) & 1 & \textbf{1.00} & \textbf{1.00}\\
\hline Zbot & 3 &(sample.exe,3), (explorer.exe,1), (kaylli.exe,1) & 1 & 0.33 & 0.60\\
\hline
\end{tabular}    
    \caption{\small{Target processes of families with multi-process propagation. $|\mathcal{M}|$ denotes the number of samples in a given family that deploys multi-process execution. $\mathcal{I}$ denotes the number of samples that have the largest intersection of targets amongst the samples in a given family, $|\mathcal{I}'|$ denotes the number of targets shared by the samples in $\mathcal{I}$ multiplied by the size of $\mathcal{I}$ and $|\mathcal{T}|$ denotes the total amount of multi-process propagations in a given family.}}
    \label{tab:FamilyTargetCounts}
\end{table*}

\begin{table*}

    \centering
    \small
    \begin{tabular}{|c|P{1.080}|c|c|c|}
    \hline
         ID & APIs & \#Total & \#Samples & \#Families \\
         \hline
1 & \texttt{OpenProcess},  \texttt{VirtualAllocEx},  \texttt{WriteProcessMemory},  \texttt{CreateRemoteThread} &  174 & 39 & 15 \\
\hline
2 & \texttt{CreateProcessW},  \texttt{VirtualAllocEx},  \texttt{WriteProcessMemory},  \texttt{SetThreadContext},  \texttt{ResumeThread} &  48 & 28 & 14 \\
\hline
3 & \texttt{CreateProcessW},  \texttt{VirtualAllocEx},  \texttt{ZwWriteVirtualMemory},  \texttt{ZwResumeThread} &  22 & 17 & 7 \\
\hline
4 & \texttt{CreateProcessInternalW},  \texttt{ZwCreateSection},  \texttt{ZwMapViewOfSection},  \texttt{ZwMapViewOfSection},  \texttt{ZwCreateThreadEx},  \texttt{ZwQueryInformationProcess},  \texttt{ZwResumeThread} &  22 & 12 & 7 \\
\hline
5 & \texttt{CreateProcessA},  \texttt{VirtualProtectEx},  \texttt{WriteProcessMemory},  \texttt{ResumeThread} &  16 & 15 & 5 \\
\hline
6 & \texttt{WinExec} &  16 & 13 & 2 \\
\hline
7 & \texttt{CreateFileA},  \texttt{WriteFile},  \texttt{WinExec} &  12 & 11 & 4 \\
\hline
8 & \texttt{CreateFileA},  \texttt{WriteFile},  \texttt{CreateProcessA} &  11 & 9 & 4 \\
\hline
9 & \texttt{ZwCreateSection},  \texttt{ZwMapViewOfSection},  \texttt{CreateProcessW},  \texttt{ZwMapViewOfSection},  \texttt{ResumeThread} &  10 & 8 & 2 \\
\hline
10 & \texttt{CreateProcessA},  \texttt{VirtualAllocEx},  \texttt{WriteProcessMemory},  \texttt{CreateRemoteThread} &  9 & 9 & 5 \\
\hline
11 & \texttt{CopyFileA},  \texttt{CreateProcessA} &  8 & 5 & 4 \\
\hline
12 & \texttt{fopen},  \texttt{\_write},  \texttt{fflush},  \texttt{CreateProcessA} &  7 & 7 & 1 \\
\hline
13 & \texttt{CreateProcessA},  \texttt{VirtualAllocEx},  \texttt{WriteProcessMemory} &  7 & 6 & 2 \\
\hline
14 & \texttt{CreateProcessW},  \texttt{ZwCreateSection},  \texttt{ZwMapViewOfSection},  \texttt{ZwMapViewOfSection},  \texttt{ZwResumeThread} &  6 & 4 & 2 \\
\hline
15 & \texttt{CreateProcessA},  \texttt{VirtualAllocEx},  \texttt{WriteProcessMemory},  \texttt{GetThreadContext},  \texttt{SetThreadContext},  \texttt{ResumeThread} &  6 & 5 & 3 \\
\hline
16 & \texttt{CreateFileW},  \texttt{WriteFile},  \texttt{CreateProcessW} &  6 & 4 & 5 \\
\hline
17 & \texttt{CreateProcessA},  \texttt{VirtualAllocEx},  \texttt{WriteProcessMemory},  \texttt{SetThreadContext},  \texttt{ResumeThread} &  5 & 3 & 5 \\
\hline
18 & \texttt{OpenProcess},  \texttt{ZwMapViewOfSection},  \texttt{ZwProtectVirtualMemory},  \texttt{ZwWriteVirtualMemory},  \texttt{CreateRemoteThread} &  4 & 4 & 1 \\
\hline
19 & \texttt{OpenProcess},  \texttt{VirtualAllocEx},  \texttt{VirtualProtectEx},  \texttt{WriteProcessMemory},  \texttt{PostMessageW} &  3 & 3 & 1 \\
\hline
20 & \texttt{GetShellWindow},  \texttt{ZwOpenProcess},  \texttt{ZwAllocateVirtualMemory},  \texttt{ZwWriteVirtualMemory},  \texttt{ZwProtectVirtualMemory},  \texttt{CreateRemoteThread} &  3 & 3 & 1 \\
\hline
21 & \texttt{CreateProcessW},  \texttt{ZwCreateSection},  \texttt{ZwMapViewOfSection},  \texttt{ZwCreateSection},  \texttt{ZwMapViewOfSection} &  3 & 3 & 1 \\
\hline
22 & \texttt{CopyFileW},  \texttt{ShellExecuteW} &  3 & 1 & 3 \\
\hline
23 & \texttt{CreateFileW},  \texttt{WriteFile},  \texttt{CreateProcessW} &  3 & 3 & 1 \\
\hline
24 & \texttt{CallWindowProcA},  \texttt{CreateProcessA},  \texttt{CallWindowProcA},  \texttt{VirtualAllocEx},  \texttt{CallWindowProcA},  \texttt{WriteProcessMemory},  \texttt{CallWindowProcA},  \texttt{GetThreadContext},  \texttt{CallWindowProcA},  \texttt{SetThreadContext},  \texttt{CallWindowProcA},  \texttt{ResumeThread} &  2 & 2 & 1 \\
\hline
25 & \texttt{CreateFileW},  \texttt{WriteFile},  \texttt{CreateProcessA} &  2 & 2 & 2 \\
\hline
26 & \texttt{CreateFileA},  \texttt{WriteFile},  \texttt{ShellExecuteA} &  2 & 2 & 2 \\
\hline
27 & \texttt{CreateFileW},  \texttt{WriteFile},  \texttt{ShellExecuteExW} &  2 & 2 & 2 \\
\hline
28 & \texttt{OpenProcess},  \texttt{DuplicateHandle},  \texttt{ZwAllocateVirtualMemory},  \texttt{WriteProcessMemory},  \texttt{CreateRemoteThread} &  1 & 1 & 1 \\
\hline
29 & \texttt{CallWindowProcA},  \texttt{CreateProcessW},  \texttt{CallWindowProcA},  \texttt{VirtualAllocEx},  \texttt{CallWindowProcA},  \texttt{WriteProcessMemory},  \texttt{CallWindowProcA},  \texttt{SetThreadContext},  \texttt{CallWindowProcA},  \texttt{ResumeThread} &  1 & 1 & 1 \\
\hline
30 & \texttt{OpenProcess},  \texttt{VirtualProtectEx},  \texttt{WriteProcessMemory},  \texttt{ResumeThread} &  1 & 1 & 1 \\
\hline
31 & \texttt{OpenProcess},  \texttt{CreateRemoteThread},  \texttt{VirtualProtectEx},  \texttt{ZwWriteVirtualMemory},  \texttt{ResumeThread} &  1 & 1 & 1 \\
\hline
32 & \texttt{CreateFileW},  \texttt{WriteFile},  \texttt{CreateFileW},  \texttt{WriteFile},  \texttt{ShellExecuteW} &  1 & 1 & 1 \\
\hline
33 & \texttt{CreateFileA},  \texttt{CreateFileMappingA},  \texttt{MapViewOfFile},  \texttt{UnmapViewOfFile},  \texttt{CreateProcessA} &  1 & 1 & 1 \\
\hline
    \end{tabular}
    \caption{The multi-process signatures observed amongst all of the samples. For each signature, the table lists the APIs involved in the signature, the total number of times the signature was observed and the number of samples and families that uses it.}
    \label{tab:SignatureDistribution}
\end{table*}

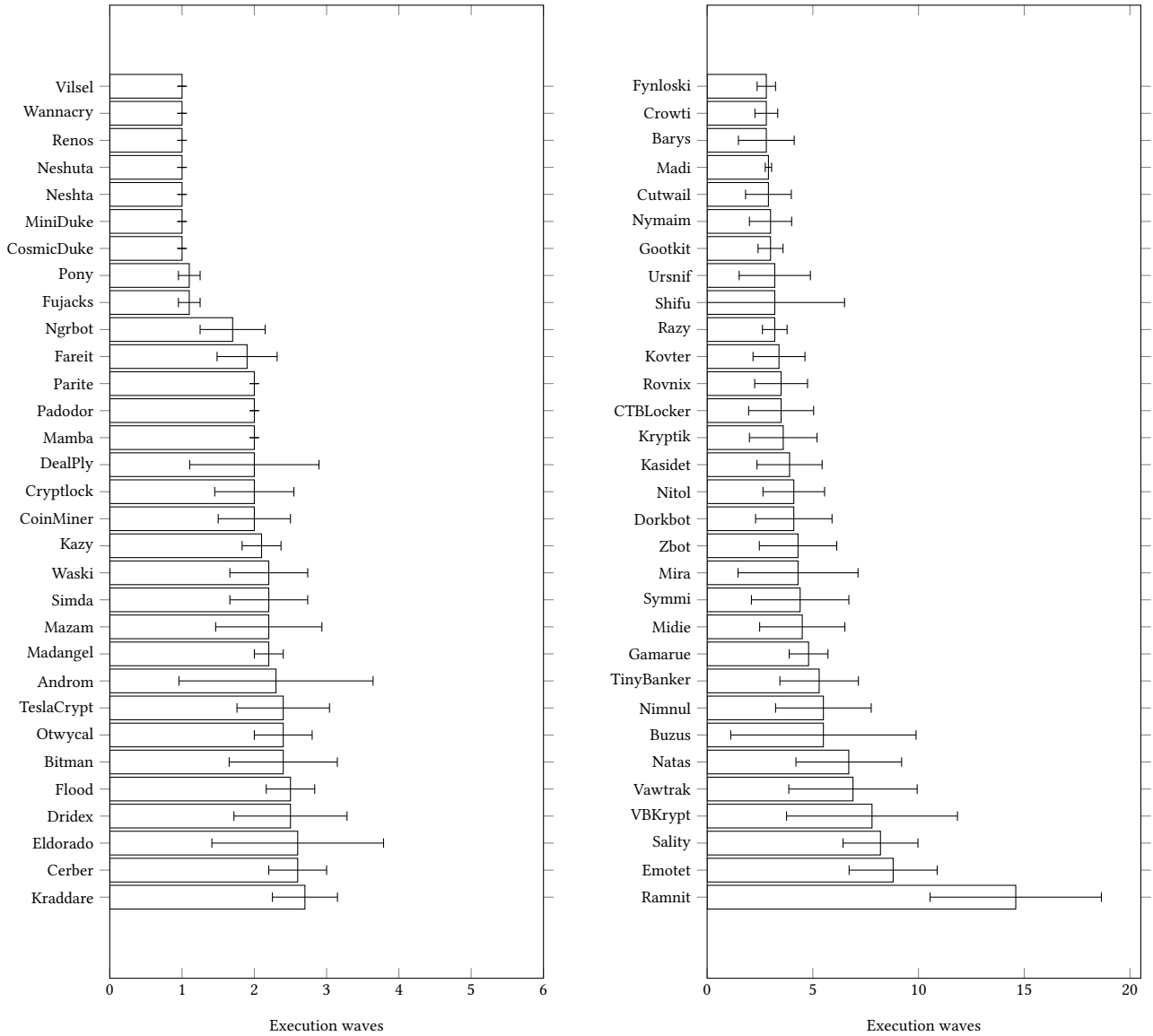
\begin{figure*}
\centering
\begin{subfigure}{.5\textwidth}
  \centering
\begin{tikzpicture}
\footnotesize
\begin{axis}[ 
    width=0.9*\textwidth,
    height=1.8*\textwidth,
xbar, xmin=0, xmax=6,
xlabel={Execution waves},
area style,
symbolic y coords={%
Kraddare,
Cerber,
Eldorado,
Dridex,
Flood,
Bitman,
Otwycal,
TeslaCrypt,
Androm,
Madangel,
Mazam,
Simda,
Waski,
Kazy,
CoinMiner,
Cryptlock,
DealPly,
Mamba,
Padodor,
Parite,
Fareit,
Ngrbot,
Fujacks,
Pony,
CosmicDuke,
MiniDuke,
Neshta,
Neshuta,
Renos,
Wannacry,
Vilsel}, ytick=data]
\addplot[error bars/.cd,
x dir=both,x explicit] coordinates {
(2.700000,Kraddare) +- (0.450000, 0.450000)
(2.600000,Cerber) +- (0.400000, 0.400000)
(2.600000,Eldorado) +- (1.187434, 1.187434)
(2.500000,Dridex) +- (0.782624, 0.782624)
(2.500000,Flood) +- (0.335410, 0.335410)
(2.400000,Bitman) +- (0.748331, 0.748331)
(2.400000,Otwycal) +- (0.400000, 0.400000)
(2.400000,TeslaCrypt) +- (0.640312, 0.640312)
(2.300000,Androm) +- (1.342572, 1.342572)
(2.200000,Madangel) +- (0.200000, 0.200000)
(2.200000,Mazam) +- (0.734847, 0.734847)
(2.200000,Simda) +- (0.538516, 0.538516)
(2.200000,Waski) +- (0.538516, 0.538516)
(2.100000,Kazy) +- (0.269258, 0.269258)
(2.000000,CoinMiner) +- (0.500000, 0.500000)
(2.000000,Cryptlock) +- (0.547723, 0.547723)
(2.000000,DealPly) +- (0.894427, 0.894427)
(2.000000,Mamba) +- (0.000000, 0.000000)
(2.000000,Padodor) +- (0.000000, 0.000000)
(2.000000,Parite) +- (0.000000, 0.000000)
(1.900000,Fareit) +- (0.415331, 0.415331)
(1.700000,Ngrbot) +- (0.450000, 0.450000)
(1.100000,Fujacks) +- (0.150000, 0.150000)
(1.100000,Pony) +- (0.150000, 0.150000)
(1.000000,CosmicDuke) +- (0.000000, 0.000000)
(1.000000,MiniDuke) +- (0.000000, 0.000000)
(1.000000,Neshta) +- (0.000000, 0.000000)
(1.000000,Neshuta) +- (0.000000, 0.000000)
(1.000000,Renos) +- (0.000000, 0.000000)
(1.000000,Wannacry) +- (0.000000, 0.000000)
(1.000000,Vilsel) +- (0.000000, 0.000000)
    };
\end{axis}
\end{tikzpicture} %width=6cm,height=7.59cm
  \label{fig:sub1}
\end{subfigure}%
\begin{subfigure}{.5\textwidth}
  \centering
\begin{tikzpicture}
\footnotesize
\begin{axis}[ 
    width=0.9*\textwidth,
    height=1.8*\textwidth,
xbar, xmin=0,
xlabel={Execution waves},
area style,
symbolic y coords={%
Ramnit,
Emotet,
Sality,
VBKrypt,
Vawtrak,
Natas,
Buzus,
Nimnul,
TinyBanker,
Gamarue,
Midie,
Symmi,
Mira,
Zbot,
Dorkbot,
Nitol,
Kasidet,
Kryptik,
CTBLocker,
Rovnix,
Kovter,
Razy,
Shifu,
Ursnif,
Gootkit,
Nymaim,
Cutwail,
Madi,
Barys,
Crowti,
Fynloski,}, ytick=data]
\addplot[error bars/.cd,
x dir=both,x explicit] coordinates {
(14.600000,Ramnit) +- (4.050926, 4.050926)
(8.800000,Emotet) +- (2.083267, 2.083267)
(8.200000,Sality) +- (1.772005, 1.772005)
(7.800000,VBKrypt) +- (4.042277, 4.042277)
(6.900000,Vawtrak) +- (3.036857, 3.036857)
(6.700000,Natas) +- (2.500500, 2.500500)
(5.500000,Buzus) +- (4.383207, 4.383207)
(5.500000,Nimnul) +- (2.261084, 2.261084)
(5.300000,TinyBanker) +- (1.858090, 1.858090)
(4.800000,Gamarue) +- (0.916515, 0.916515)
(4.500000,Midie) +- (2.015564, 2.015564)
(4.400000,Symmi) +- (2.304344, 2.304344)
(4.300000,Mira) +- (2.837693, 2.837693)
(4.300000,Zbot) +- (1.830983, 1.830983)
(4.100000,Dorkbot) +- (1.809005, 1.809005)
(4.100000,Nitol) +- (1.456880, 1.456880)
(3.900000,Kasidet) +- (1.540292, 1.540292)
(3.600000,Kryptik) +- (1.600000, 1.600000)
(3.500000,CTBLocker) +- (1.537043, 1.537043)
(3.500000,Rovnix) +- (1.250000, 1.250000)
(3.400000,Kovter) +- (1.228821, 1.228821)
(3.200000,Razy) +- (0.583095, 0.583095)
(3.200000,Shifu) +- (3.300000, 3.300000)
(3.200000,Ursnif) +- (1.685230, 1.685230)
(3.000000,Gootkit) +- (0.591608, 0.591608)
(3.000000,Nymaim) +- (1.000000, 1.000000)
(2.900000,Cutwail) +- (1.082820, 1.082820)
(2.900000,Madi) +- (0.150000, 0.150000)
(2.800000,Barys) +- (1.319091, 1.319091)
(2.800000,Crowti) +- (0.538516, 0.538516)
(2.800000,Fynloski) +- (0.435890, 0.435890)
    };
\end{axis}
\end{tikzpicture} %width=6cm,height=7.59cm
  \label{fig:sub2}
\end{subfigure}
\centering
\caption{Average number and standard deviation of execution waves per family. The plots are sorted by average starting from the top left, i.e. notice the difference in x-axis values between the two plots.} 
\label{fig:PerFamilyExecutionWaves}  
\end{figure*}

\end{document}

%% file: AlgorithmStyle.tex
\usepackage{algorithm2e}
\usepackage{xcolor}

\SetAlFnt{\footnotesize}

\SetAlCapSty{xAlCapSty}

\SetCommentSty{xCommentSty}

\SetNlSty{mynlfont}{}{} 

\LinesNumbered

\SetSideCommentRight

\DontPrintSemicolon

\RestyleAlgo{algoruled}

\definecolor{mGreen}{rgb}{0,0.6,0}
\definecolor{mGray}{rgb}{0.5,0.5,0.5}
\definecolor{mPurple}{rgb}{0.58,0,0.82}
\definecolor{backgroundColour}{rgb}{0.95,0.95,0.92}

\lstdefinestyle{CStyle}{
    backgroundcolor=\color{backgroundColour},   
    commentstyle=\color{mGreen},
    keywordstyle=\color{magenta},
    numberstyle=\tiny\color{mGray},
    stringstyle=\color{mPurple},
    basicstyle=\footnotesize,
    breakatwhitespace=false,         
    breaklines=true,                 
    captionpos=b,                    
    keepspaces=true,                 
    numbers=left,                    
    numbersep=5pt,                  
    showspaces=false,                
    showstringspaces=false,
    showtabs=false,                  
    tabsize=2,
    language=C
}

%% file: mytikz.tex
\usepackage[utf8]{inputenc}

\usepackage{tikz}
\usetikzlibrary{shapes,shadows}
\usetikzlibrary{arrows,automata}
\usetikzlibrary{positioning}
\usetikzlibrary{calc,positioning}
\usetikzlibrary{decorations.markings}

\makeatletter
\tikzset{%
  remember picture with id/.style={%
    remember picture,
    overlay,
    save picture id=#1,
  },
  save picture id/.code={%
    \edef\pgf@temp{#1}%
    \immediate\write\pgfutil@auxout{%
      \noexpand\savepointas{\pgf@temp}{\pgfpictureid}}%
  },
  if picture id/.code args={#1#2#3}{%
    \@ifundefined{save@pt@#1}{%
      \pgfkeysalso{#3}%
    }{
      \pgfkeysalso{#2}%
    }
  }
}

\def\savepointas#1#2{%
  \expandafter\gdef\csname save@pt@#1\endcsname{#2}%
}

\def\tmk@labeldef#1,#2\@nil{%
  \def\tmk@label{#1}%
  \def\tmk@def{#2}%
}

\tikzdeclarecoordinatesystem{pic}{%
  \pgfutil@in@,{#1}%
  \ifpgfutil@in@%
    \tmk@labeldef#1\@nil
  \else
    \tmk@labeldef#1,(0pt,0pt)\@nil
  \fi
  \@ifundefined{save@pt@\tmk@label}{%
    \tikz@scan@one@point\pgfutil@firstofone\tmk@def
  }{%
  \pgfsys@getposition{\csname save@pt@\tmk@label\endcsname}\save@orig@pic%
  \pgfsys@getposition{\pgfpictureid}\save@this@pic%
  \pgf@process{\pgfpointorigin\save@this@pic}%
  \pgf@xa=\pgf@x
  \pgf@ya=\pgf@y
  \pgf@process{\pgfpointorigin\save@orig@pic}%
  \advance\pgf@x by -\pgf@xa
  \advance\pgf@y by -\pgf@ya
  }%
}

\makeatother
\usepackage{caption}

\usepackage{adjustbox}

\usetikzlibrary{positioning}
\usetikzlibrary{fit,backgrounds}

\tikzset{
    state/.style={
           rectangle,
           rounded corners,
           draw=black, very thick,
           minimum height=2em,
           inner sep=2pt,
           },
    propstate/.style={
           rectangle,
           draw=black, very thick,
           minimum height=2em,
           inner sep=2pt,
           },
}

\usepackage{color, colortbl}
\definecolor{Gray}{gray}{0.9}
\definecolor{lightblue}{rgb}{0.93,0.95,1.0}
\definecolor{myyellow}{rgb}{0.95, 0.95, 0}
\definecolor{navy_blue}{rgb}{0.7, 0.7, 0.90}
\definecolor{red}{rgb}{0.9, 0.6, 0.4}
\definecolor{pale_green}{rgb}{0.75, 0.96, 0.75}

\pgfdeclarelayer{background}
\pgfdeclarelayer{foreground}
\pgfsetlayers{background,main,foreground}

\tikzstyle{plain}=[draw, fill=gray!00, text width=10em, text centered, minimum height=1.5em]
\tikzstyle{thickplain}=[draw, fill=gray!00, text width=5em, text centered, minimum height=2.5em]
\tikzstyle{malplain}=[draw, fill=gray!00, text width=4em, text centered, minimum height=2.5em]

\tikzstyle{textplain}=[text width=10em, text centered]

\tikzstyle{addressbox}=[draw, fill=gray!00, text width=5em, text centered, minimum height=1.5em]

\tikzstyle{sensor}=[draw, fill=blue!20, text width=5em, 
    text centered, minimum height=2.5em,drop shadow]
\tikzstyle{ann} = [above, text width=5em, text centered]
\tikzstyle{wa} = [sensor, text width=10em, fill=red!20, 
    minimum height=6em, rounded corners, drop shadow]
\tikzstyle{sc} = [sensor, text width=13em, fill=red!20, 
    minimum height=10em, rounded corners, drop shadow]
\tikzstyle{wave} = [draw, fill=gray!30, text width=4em, 
    text centered, minimum height=2.5em,drop shadow, rounded corners, drop shadow]
\tikzstyle{wave2} = [draw, fill=gray!30, text width=7em, 
    text centered, minimum height=2.5em,drop shadow, rounded corners, drop shadow]

\tikzstyle{code} = [draw, fill=gray!20, %text width=13em, 
    minimum height=2.5em,drop shadow]
\tikzstyle{codereuse} = [draw, fill=black!75, text width = 5em,
	text centered, minimum height=2.5em, drop shadow,text=white]

\tikzstyle{docs} = [draw, minimum height=4em, minimum width=3em, 
                fill=white, 
                double copy shadow={shadow xshift=-4pt, 
                             shadow yshift=-4pt, fill=white, draw}]

%% file: sample-sigconf.bbl
%%% -*-BibTeX-*-
%%% Do NOT edit. File created by BibTeX with style
%%% ACM-Reference-Format-Journals [18-Jan-2012].

\begin{thebibliography}{36}

%%% ====================================================================
%%% NOTE TO THE USER: you can override these defaults by providing
%%% customized versions of any of these macros before the \bibliography
%%% command.  Each of them MUST provide its own final punctuation,
%%% except for \shownote{}, \showDOI{}, and \showURL{}.  The latter two
%%% do not use final punctuation, in order to avoid confusing it with
%%% the Web address.
%%%
%%% To suppress output of a particular field, define its macro to expand
%%% to an empty string, or better, \unskip, like this:
%%%
%%% \newcommand{\showDOI}[1]{\unskip}   % LaTeX syntax
%%%
%%% \def \showDOI #1{\unskip}           % plain TeX syntax
%%%
%%% ====================================================================

\ifx \showCODEN    \undefined \def \showCODEN     #1{\unskip}     \fi
\ifx \showDOI      \undefined \def \showDOI       #1{#1}\fi
\ifx \showISBNx    \undefined \def \showISBNx     #1{\unskip}     \fi
\ifx \showISBNxiii \undefined \def \showISBNxiii  #1{\unskip}     \fi
\ifx \showISSN     \undefined \def \showISSN      #1{\unskip}     \fi
\ifx \showLCCN     \undefined \def \showLCCN      #1{\unskip}     \fi
\ifx \shownote     \undefined \def \shownote      #1{#1}          \fi
\ifx \showarticletitle \undefined \def \showarticletitle #1{#1}   \fi
\ifx \showURL      \undefined \def \showURL       {\relax}        \fi
% The following commands are used for tagged output and should be
% invisible to TeX
\providecommand\bibfield[2]{#2}
\providecommand\bibinfo[2]{#2}
\providecommand\natexlab[1]{#1}
\providecommand\showeprint[2][]{arXiv:#2}

\bibitem[\protect\citeauthoryear{Bacs, Vermeulen, Slowinska, and Bos}{Bacs
  et~al\mbox{.}}{2013}]%
        {10.1007/978-3-642-37300-8_9}
\bibfield{author}{\bibinfo{person}{Andrei Bacs}, \bibinfo{person}{Remco
  Vermeulen}, \bibinfo{person}{Asia Slowinska}, {and} \bibinfo{person}{Herbert
  Bos}.} \bibinfo{year}{2013}\natexlab{}.
\newblock \showarticletitle{System-Level Support for Intrusion Recovery}. In
  \bibinfo{booktitle}{\emph{Detection of Intrusions and Malware, and
  Vulnerability Assessment}}, \bibfield{editor}{\bibinfo{person}{Ulrich
  Flegel}, \bibinfo{person}{Evangelos Markatos}, {and} \bibinfo{person}{William
  Robertson}} (Eds.). \bibinfo{publisher}{Springer Berlin Heidelberg},
  \bibinfo{address}{Berlin, Heidelberg}, \bibinfo{pages}{144--163}.
\newblock
\showISBNx{978-3-642-37300-8}


\bibitem[\protect\citeauthoryear{Barabosch, Bergmann, Dombeck, and
  Padilla}{Barabosch et~al\mbox{.}}{2017}]%
        {DBLP:conf/dimva/BaraboschBDP17}
\bibfield{author}{\bibinfo{person}{Thomas Barabosch}, \bibinfo{person}{Niklas
  Bergmann}, \bibinfo{person}{Adrian Dombeck}, {and} \bibinfo{person}{Elmar
  Padilla}.} \bibinfo{year}{2017}\natexlab{}.
\newblock \showarticletitle{Quincy: Detecting Host-Based Code Injection Attacks
  in Memory Dumps}. In \bibinfo{booktitle}{\emph{Detection of Intrusions and
  Malware, and Vulnerability Assessment - 14th International Conference,
  {DIMVA} 2017, Bonn, Germany, July 6-7, 2017, Proceedings}}.
  \bibinfo{pages}{209--229}.
\newblock
\urldef\tempurl%
\url{https://doi.org/10.1007/978-3-319-60876-1\_10}
\showDOI{\tempurl}


\bibitem[\protect\citeauthoryear{Barabosch, Eschweiler, and
  Gerhards{-}Padilla}{Barabosch et~al\mbox{.}}{2014}]%
        {DBLP:conf/dimva/BaraboschEG14}
\bibfield{author}{\bibinfo{person}{Thomas Barabosch},
  \bibinfo{person}{Sebastian Eschweiler}, {and} \bibinfo{person}{Elmar
  Gerhards{-}Padilla}.} \bibinfo{year}{2014}\natexlab{}.
\newblock \showarticletitle{Bee Master: Detecting Host-Based Code Injection
  Attacks}. In \bibinfo{booktitle}{\emph{Detection of Intrusions and Malware,
  and Vulnerability Assessment - 11th International Conference, {DIMVA} 2014,
  Egham, UK, July 10-11, 2014. Proceedings}}. \bibinfo{pages}{235--254}.
\newblock
\urldef\tempurl%
\url{https://doi.org/10.1007/978-3-319-08509-8\_13}
\showDOI{\tempurl}


\bibitem[\protect\citeauthoryear{Barabosch and Gerhards{-}Padilla}{Barabosch
  and Gerhards{-}Padilla}{2014}]%
        {DBLP:conf/malware/BaraboschG14}
\bibfield{author}{\bibinfo{person}{Thomas Barabosch} {and}
  \bibinfo{person}{Elmar Gerhards{-}Padilla}.} \bibinfo{year}{2014}\natexlab{}.
\newblock \showarticletitle{Host-based code injection attacks: {A} popular
  technique used by malware}. In \bibinfo{booktitle}{\emph{9th International
  Conference on Malicious and Unwanted Software: The Americas {MALWARE} 2014,
  Fajardo, PR, USA, October 28-30, 2014}}. \bibinfo{pages}{8--17}.
\newblock
\urldef\tempurl%
\url{https://doi.org/10.1109/MALWARE.2014.6999410}
\showDOI{\tempurl}


\bibitem[\protect\citeauthoryear{Bayer, Habibi, Balzarotti, and Kirda}{Bayer
  et~al\mbox{.}}{2009}]%
        {DBLP:conf/leet/BayerHBK09}
\bibfield{author}{\bibinfo{person}{Ulrich Bayer}, \bibinfo{person}{Imam
  Habibi}, \bibinfo{person}{Davide Balzarotti}, {and} \bibinfo{person}{Engin
  Kirda}.} \bibinfo{year}{2009}\natexlab{}.
\newblock \showarticletitle{A View on Current Malware Behaviors}. In
  \bibinfo{booktitle}{\emph{2nd {USENIX} Workshop on Large-Scale Exploits and
  Emergent Threats, {LEET} '09, Boston, MA, USA, April 21, 2009}}.
\newblock
\urldef\tempurl%
\url{https://www.usenix.org/conference/leet-09/view-current-malware-behaviors}
\showURL{%
\tempurl}


\bibitem[\protect\citeauthoryear{Bayer, Moser, Kruegel, and Kirda}{Bayer
  et~al\mbox{.}}{2006}]%
        {Bayer2006}
\bibfield{author}{\bibinfo{person}{Ulrich Bayer}, \bibinfo{person}{Andreas
  Moser}, \bibinfo{person}{Christopher Kruegel}, {and} \bibinfo{person}{Engin
  Kirda}.} \bibinfo{year}{2006}\natexlab{}.
\newblock \showarticletitle{Dynamic Analysis of Malicious Code}.
\newblock \bibinfo{journal}{\emph{Journal in Computer Virology}}
  \bibinfo{volume}{2}, \bibinfo{number}{1} (\bibinfo{date}{01 Aug}
  \bibinfo{year}{2006}), \bibinfo{pages}{67--77}.
\newblock
\showISSN{1772-9904}
\urldef\tempurl%
\url{https://doi.org/10.1007/s11416-006-0012-2}
\showDOI{\tempurl}


\bibitem[\protect\citeauthoryear{Baz and Safran}{Baz and Safran}{2017}]%
        {DridexV4Atombinging}
\bibfield{author}{\bibinfo{person}{Magal Baz} {and} \bibinfo{person}{Or
  Safran}.} \bibinfo{year}{2017}\natexlab{}.
\newblock \bibinfo{title}{Dridex’s Cold War: Enter AtomBombing}.
\newblock
\newblock


\bibitem[\protect\citeauthoryear{Bellard}{Bellard}{2005}]%
        {Bellard:2005:QFP:1247360.1247401}
\bibfield{author}{\bibinfo{person}{Fabrice Bellard}.}
  \bibinfo{year}{2005}\natexlab{}.
\newblock \showarticletitle{QEMU, a Fast and Portable Dynamic Translator}. In
  \bibinfo{booktitle}{\emph{Proceedings of the Annual Conference on USENIX
  Annual Technical Conference}} \emph{(\bibinfo{series}{ATEC '05})}.
  \bibinfo{publisher}{USENIX Association}, \bibinfo{address}{Berkeley, CA,
  USA}, \bibinfo{pages}{41--41}.
\newblock
\urldef\tempurl%
\url{http://dl.acm.org/citation.cfm?id=1247360.1247401}
\showURL{%
\tempurl}


\bibitem[\protect\citeauthoryear{Bonfante, Fernandez, Marion, Rouxel, Sabatier,
  and Thierry}{Bonfante et~al\mbox{.}}{2015}]%
        {Bonfante:2015:CMS:2810103.2813627}
\bibfield{author}{\bibinfo{person}{Guillaume Bonfante}, \bibinfo{person}{Jose
  Fernandez}, \bibinfo{person}{Jean-Yves Marion}, \bibinfo{person}{Benjamin
  Rouxel}, \bibinfo{person}{Fabrice Sabatier}, {and}
  \bibinfo{person}{Aur{\'e}lien Thierry}.} \bibinfo{year}{2015}\natexlab{}.
\newblock \showarticletitle{CoDisasm: Medium Scale Concatic Disassembly of
  Self-Modifying Binaries with Overlapping Instructions}. In
  \bibinfo{booktitle}{\emph{Proceedings of the 22Nd ACM SIGSAC Conference on
  Computer and Communications Security}} \emph{(\bibinfo{series}{CCS '15})}.
  \bibinfo{publisher}{ACM}, \bibinfo{address}{New York, NY, USA},
  \bibinfo{pages}{745--756}.
\newblock
\showISBNx{978-1-4503-3832-5}
\urldef\tempurl%
\url{https://doi.org/10.1145/2810103.2813627}
\showDOI{\tempurl}


\bibitem[\protect\citeauthoryear{Cozzi, Graziano, Fratantonio, and
  Balzarotti}{Cozzi et~al\mbox{.}}{[n.d.]}]%
        {UnderstandingLinuxMalware}
\bibfield{author}{\bibinfo{person}{E. Cozzi}, \bibinfo{person}{M. Graziano},
  \bibinfo{person}{Y. Fratantonio}, {and} \bibinfo{person}{D. Balzarotti}.}
  \bibinfo{year}{[n.d.]}\natexlab{}.
\newblock \showarticletitle{Understanding Linux Malware}. In
  \bibinfo{booktitle}{\emph{2018 IEEE Symposium on Security and Privacy (SP)}},
  Vol.~\bibinfo{volume}{00}. \bibinfo{pages}{870--884}.
\newblock
\showISSN{2375-1207}
\urldef\tempurl%
\url{https://doi.org/10.1109/SP.2018.00054}
\showDOI{\tempurl}


\bibitem[\protect\citeauthoryear{{C}ozzi, {G}raziano, {F}ratantonio, and
  {B}alzarotti}{{C}ozzi et~al\mbox{.}}{2018}]%
        {EURECOM+5489}
\bibfield{author}{\bibinfo{person}{{E}manuele {C}ozzi},
  \bibinfo{person}{{M}ariano {G}raziano}, \bibinfo{person}{{Y}anick
  {F}ratantonio}, {and} \bibinfo{person}{{D}avide {B}alzarotti}.}
  \bibinfo{year}{2018}\natexlab{}.
\newblock \showarticletitle{{U}nderstanding {L}inux malware}. In
  \bibinfo{booktitle}{\emph{{S}\&{P} 2018, 39th {IEEE} {S}ymposium on
  {S}ecurity and {P}rivacy, {M}ay 21-23, 2018, {S}an {F}rancisco, {CA},
  {USA}}}. \bibinfo{address}{{S}an {F}rancisco, {UNITED} {STATES}}.
\newblock
\urldef\tempurl%
\url{http://www.eurecom.fr/publication/5489}
\showURL{%
\tempurl}


\bibitem[\protect\citeauthoryear{Dinaburg, Royal, Sharif, and Lee}{Dinaburg
  et~al\mbox{.}}{2008}]%
        {Dinaburg:2008:EMA:1455770.1455779}
\bibfield{author}{\bibinfo{person}{Artem Dinaburg}, \bibinfo{person}{Paul
  Royal}, \bibinfo{person}{Monirul Sharif}, {and} \bibinfo{person}{Wenke Lee}.}
  \bibinfo{year}{2008}\natexlab{}.
\newblock \showarticletitle{Ether: Malware Analysis via Hardware Virtualization
  Extensions}. In \bibinfo{booktitle}{\emph{Proceedings of the 15th ACM
  Conference on Computer and Communications Security}}
  \emph{(\bibinfo{series}{CCS '08})}. \bibinfo{publisher}{ACM},
  \bibinfo{address}{New York, NY, USA}, \bibinfo{pages}{51--62}.
\newblock
\showISBNx{978-1-59593-810-7}
\urldef\tempurl%
\url{https://doi.org/10.1145/1455770.1455779}
\showDOI{\tempurl}


\bibitem[\protect\citeauthoryear{Dolan-Gavitt, Hodosh, Hulin, Leek, and
  Whelan}{Dolan-Gavitt et~al\mbox{.}}{2015}]%
        {Dolan-Gavitt:2015:RRE:2843859.2843867}
\bibfield{author}{\bibinfo{person}{Brendan Dolan-Gavitt}, \bibinfo{person}{Josh
  Hodosh}, \bibinfo{person}{Patrick Hulin}, \bibinfo{person}{Tim Leek}, {and}
  \bibinfo{person}{Ryan Whelan}.} \bibinfo{year}{2015}\natexlab{}.
\newblock \showarticletitle{Repeatable Reverse Engineering with PANDA}. In
  \bibinfo{booktitle}{\emph{Proceedings of the 5th Program Protection and
  Reverse Engineering Workshop}} \emph{(\bibinfo{series}{PPREW-5})}.
  \bibinfo{publisher}{ACM}, \bibinfo{address}{New York, NY, USA}, Article
  \bibinfo{articleno}{4}, \bibinfo{numpages}{11}~pages.
\newblock
\showISBNx{978-1-4503-3642-0}
\urldef\tempurl%
\url{https://doi.org/10.1145/2843859.2843867}
\showDOI{\tempurl}


\bibitem[\protect\citeauthoryear{Egele, Scholte, Kirda, and Kruegel}{Egele
  et~al\mbox{.}}{2008}]%
        {Egele:2008:SAD:2089125.2089126}
\bibfield{author}{\bibinfo{person}{Manuel Egele}, \bibinfo{person}{Theodoor
  Scholte}, \bibinfo{person}{Engin Kirda}, {and} \bibinfo{person}{Christopher
  Kruegel}.} \bibinfo{year}{2008}\natexlab{}.
\newblock \showarticletitle{A Survey on Automated Dynamic Malware-analysis
  Techniques and Tools}.
\newblock \bibinfo{journal}{\emph{ACM Comput. Surv.}} \bibinfo{volume}{44},
  \bibinfo{number}{2}, Article \bibinfo{articleno}{6} (\bibinfo{date}{March}
  \bibinfo{year}{2008}), \bibinfo{numpages}{42}~pages.
\newblock
\showISSN{0360-0300}
\urldef\tempurl%
\url{https://doi.org/10.1145/2089125.2089126}
\showDOI{\tempurl}


\bibitem[\protect\citeauthoryear{Felt, Finifter, Chin, Hanna, and Wagner}{Felt
  et~al\mbox{.}}{2011}]%
        {Felt:2011:SMM:2046614.2046618}
\bibfield{author}{\bibinfo{person}{Adrienne~Porter Felt},
  \bibinfo{person}{Matthew Finifter}, \bibinfo{person}{Erika Chin},
  \bibinfo{person}{Steve Hanna}, {and} \bibinfo{person}{David Wagner}.}
  \bibinfo{year}{2011}\natexlab{}.
\newblock \showarticletitle{A Survey of Mobile Malware in the Wild}. In
  \bibinfo{booktitle}{\emph{Proceedings of the 1st ACM Workshop on Security and
  Privacy in Smartphones and Mobile Devices}} \emph{(\bibinfo{series}{SPSM
  '11})}. \bibinfo{publisher}{ACM}, \bibinfo{address}{New York, NY, USA},
  \bibinfo{pages}{3--14}.
\newblock
\showISBNx{978-1-4503-1000-0}
\urldef\tempurl%
\url{https://doi.org/10.1145/2046614.2046618}
\showDOI{\tempurl}


\bibitem[\protect\citeauthoryear{HASHEREZADE}{HASHEREZADE}{2016}]%
        {HASHEREZADECodeInjection}
\bibfield{author}{\bibinfo{person}{HASHEREZADE}.}
  \bibinfo{year}{2016}\natexlab{}.
\newblock \bibinfo{howpublished}{\url{https://github.com/hasherezade/demos}}.
\newblock


\bibitem[\protect\citeauthoryear{Henderson, Yan, Hu, Prakash, Yin, and
  McCamant}{Henderson et~al\mbox{.}}{2017}]%
        {Henderson:2017:DPW:3057931.3057958}
\bibfield{author}{\bibinfo{person}{Andrew Henderson},
  \bibinfo{person}{Lok-Kwong Yan}, \bibinfo{person}{Xunchao Hu},
  \bibinfo{person}{Aravind Prakash}, \bibinfo{person}{Heng Yin}, {and}
  \bibinfo{person}{Stephen McCamant}.} \bibinfo{year}{2017}\natexlab{}.
\newblock \showarticletitle{DECAF: A Platform-Neutral Whole-System Dynamic
  Binary Analysis Platform}.
\newblock \bibinfo{journal}{\emph{IEEE Trans. Softw. Eng.}}
  \bibinfo{volume}{43}, \bibinfo{number}{2} (\bibinfo{date}{Feb.}
  \bibinfo{year}{2017}), \bibinfo{pages}{164--184}.
\newblock
\showISSN{0098-5589}
\urldef\tempurl%
\url{https://doi.org/10.1109/TSE.2016.2589242}
\showDOI{\tempurl}


\bibitem[\protect\citeauthoryear{Hosseini}{Hosseini}{2017}]%
        {endgame_blog}
\bibfield{author}{\bibinfo{person}{Ashkan Hosseini}.}
  \bibinfo{year}{2017}\natexlab{}.
\newblock \bibinfo{title}{Ten Process Injection Techniques: A Technical Survey
  Of Common And Trending Process Injection Techniques}.
\newblock
  \bibinfo{howpublished}{\url{https://www.endgame.com/blog/technical-blog/ten-process-injection-techniques-technical-survey-common-and-trending-process}}.
\newblock


\bibitem[\protect\citeauthoryear{Hungenberg and Eckert}{Hungenberg and
  Eckert}{2018}]%
        {InetSim}
\bibfield{author}{\bibinfo{person}{Thomas Hungenberg} {and}
  \bibinfo{person}{Matthias Eckert}.} \bibinfo{year}{2018}\natexlab{}.
\newblock \bibinfo{howpublished}{\url{http://www.inetsim.org/}}.
\newblock


\bibitem[\protect\citeauthoryear{Ispoglou and Payer}{Ispoglou and
  Payer}{2016}]%
        {198415}
\bibfield{author}{\bibinfo{person}{Kyriakos~K. Ispoglou} {and}
  \bibinfo{person}{Mathias Payer}.} \bibinfo{year}{2016}\natexlab{}.
\newblock \showarticletitle{malWASH: Washing Malware to Evade Dynamic
  Analysis}. In \bibinfo{booktitle}{\emph{10th {USENIX} Workshop on Offensive
  Technologies ({WOOT} 16)}}. \bibinfo{publisher}{{USENIX} Association},
  \bibinfo{address}{Austin, TX}.
\newblock
\urldef\tempurl%
\url{https://www.usenix.org/conference/woot16/workshop-program/presentation/ispoglou}
\showURL{%
\tempurl}


\bibitem[\protect\citeauthoryear{Kang, Poosankam, and Yin}{Kang
  et~al\mbox{.}}{2007}]%
        {Kang:2007:RHC:1314389.1314399}
\bibfield{author}{\bibinfo{person}{Min~Gyung Kang}, \bibinfo{person}{Pongsin
  Poosankam}, {and} \bibinfo{person}{Heng Yin}.}
  \bibinfo{year}{2007}\natexlab{}.
\newblock \showarticletitle{Renovo: A Hidden Code Extractor for Packed
  Executables}. In \bibinfo{booktitle}{\emph{Proceedings of the 2007 ACM
  Workshop on Recurring Malcode}} \emph{(\bibinfo{series}{WORM '07})}.
  \bibinfo{publisher}{ACM}, \bibinfo{address}{New York, NY, USA},
  \bibinfo{pages}{46--53}.
\newblock
\showISBNx{978-1-59593-886-2}
\urldef\tempurl%
\url{https://doi.org/10.1145/1314389.1314399}
\showDOI{\tempurl}


\bibitem[\protect\citeauthoryear{Kawakoya, Shioji, Iwamura, and
  Miyoshi}{Kawakoya et~al\mbox{.}}{2019}]%
        {YuheiKawakoya2019}
\bibfield{author}{\bibinfo{person}{Yuhei Kawakoya}, \bibinfo{person}{Eitaro
  Shioji}, \bibinfo{person}{Makoto Iwamura}, {and} \bibinfo{person}{Jun
  Miyoshi}.} \bibinfo{year}{2019}\natexlab{}.
\newblock \showarticletitle{API Chaser: Taint-Assisted Sandbox for Evasive
  Malware Analysis}.
\newblock \bibinfo{journal}{\emph{Journal of Information Processing}}
  \bibinfo{volume}{27} (\bibinfo{year}{2019}), \bibinfo{pages}{297--314}.
\newblock
\urldef\tempurl%
\url{https://doi.org/10.2197/ipsjjip.27.297}
\showDOI{\tempurl}


\bibitem[\protect\citeauthoryear{Korczynski}{Korczynski}{2016}]%
        {DBLP:conf/malware/Korczynski16}
\bibfield{author}{\bibinfo{person}{David Korczynski}.}
  \bibinfo{year}{2016}\natexlab{}.
\newblock \showarticletitle{RePEconstruct: reconstructing binaries with
  self-modifying code and import address table destruction}. In
  \bibinfo{booktitle}{\emph{IEEE 11th International Conference on Malicious and
  Unwanted Software, {MALWARE} 2016, Fajardo, PR, USA, October 18-21, 2016}}.
  \bibinfo{publisher}{{IEEE} Computer Society}, \bibinfo{pages}{31--38}.
\newblock
\showISBNx{978-1-5090-4542-6}
\urldef\tempurl%
\url{https://doi.org/10.1109/MALWARE.2016.7888727}
\showDOI{\tempurl}


\bibitem[\protect\citeauthoryear{{Korczynski}}{{Korczynski}}{2019}]%
        {2019arXiv190809204K}
\bibfield{author}{\bibinfo{person}{David {Korczynski}}.}
  \bibinfo{year}{2019}\natexlab{}.
\newblock \showarticletitle{{Precise system-wide concatic malware unpacking}}.
\newblock \bibinfo{journal}{\emph{arXiv e-prints}}, Article
  \bibinfo{articleno}{arXiv:1908.09204} (\bibinfo{date}{Aug}
  \bibinfo{year}{2019}), \bibinfo{numpages}{arXiv:1908.09204}~pages.
\newblock
\showeprint[arxiv]{cs.CR/1908.09204}


\bibitem[\protect\citeauthoryear{Korczynski and Yin}{Korczynski and
  Yin}{2017}]%
        {DKOR}
\bibfield{author}{\bibinfo{person}{David Korczynski} {and}
  \bibinfo{person}{Heng Yin}.} \bibinfo{year}{2017}\natexlab{}.
\newblock \showarticletitle{Capturing Malware Propagations with Code Injections
  and Code-Reuse Attacks}. In \bibinfo{booktitle}{\emph{Proceedings of the 2017
  {ACM} {SIGSAC} Conference on Computer and Communications Security, {CCS}
  2017, Dallas, TX, USA, October 30 - November 03, 2017}},
  \bibfield{editor}{\bibinfo{person}{Bhavani~M. Thuraisingham},
  \bibinfo{person}{David Evans}, \bibinfo{person}{Tal Malkin}, {and}
  \bibinfo{person}{Dongyan Xu}} (Eds.). \bibinfo{publisher}{{ACM}},
  \bibinfo{pages}{1691--1708}.
\newblock
\showISBNx{978-1-4503-4946-8}
\urldef\tempurl%
\url{https://doi.org/10.1145/3133956.3134099}
\showDOI{\tempurl}


\bibitem[\protect\citeauthoryear{Networks}{Networks}{2013}]%
        {PaloAltoReport}
\bibfield{author}{\bibinfo{person}{PaloAlto Networks}.}
  \bibinfo{year}{2013}\natexlab{}.
\newblock \bibinfo{title}{The Modern Malware Review}.
\newblock
\newblock


\bibitem[\protect\citeauthoryear{Pasquale}{Pasquale}{2017}]%
        {Injectopi}
\bibfield{author}{\bibinfo{person}{Giulio~De Pasquale}.}
  \bibinfo{year}{2017}\natexlab{}.
\newblock \bibinfo{howpublished}{\url{https://github.com/peperunas/injectopi}}.
\newblock


\bibitem[\protect\citeauthoryear{Plohmann, Clauß, and Padilla}{Plohmann
  et~al\mbox{.}}{2017}]%
        {CybIN}
\bibfield{author}{\bibinfo{person}{Daniel Plohmann}, \bibinfo{person}{Martin
  Clauß}, {and} \bibinfo{person}{Elmar Padilla}.}
  \bibinfo{year}{2017}\natexlab{}.
\newblock \showarticletitle{Malpedia: A Collaborative Effort to Inventorize the
  Malware Landscape}.
\newblock \bibinfo{journal}{\emph{The Journal on Cybercrime \& Digital
  Investigations}} \bibinfo{volume}{3}, \bibinfo{number}{1}
  (\bibinfo{year}{2017}), \bibinfo{pages}{1--19}.
\newblock
\showISSN{2494-2715}
\urldef\tempurl%
\url{https://doi.org/10.18464/cybin.v3i1.17}
\showDOI{\tempurl}


\bibitem[\protect\citeauthoryear{Portokalidis, Slowinska, and Bos}{Portokalidis
  et~al\mbox{.}}{2006}]%
        {argos:eurosys06}
\bibfield{author}{\bibinfo{person}{Georgios Portokalidis},
  \bibinfo{person}{Asia Slowinska}, {and} \bibinfo{person}{Herbert Bos}.}
  \bibinfo{year}{2006}\natexlab{}.
\newblock \showarticletitle{Argos: an Emulator for Fingerprinting Zero-Day
  Attacks}. In \bibinfo{booktitle}{\emph{Proc. ACM SIGOPS EUROSYS'2006}}.
  \bibinfo{address}{Leuven, Belgium}.
\newblock


\bibitem[\protect\citeauthoryear{Sebasti{\'a}n, Rivera, Kotzias, and
  Caballero}{Sebasti{\'a}n et~al\mbox{.}}{2016}]%
        {10.1007/978-3-319-45719-2_11}
\bibfield{author}{\bibinfo{person}{Marcos Sebasti{\'a}n},
  \bibinfo{person}{Richard Rivera}, \bibinfo{person}{Platon Kotzias}, {and}
  \bibinfo{person}{Juan Caballero}.} \bibinfo{year}{2016}\natexlab{}.
\newblock \showarticletitle{AVclass: A Tool for Massive Malware Labeling}. In
  \bibinfo{booktitle}{\emph{Research in Attacks, Intrusions, and Defenses}},
  \bibfield{editor}{\bibinfo{person}{Fabian Monrose}, \bibinfo{person}{Marc
  Dacier}, \bibinfo{person}{Gregory Blanc}, {and} \bibinfo{person}{Joaquin
  Garcia-Alfaro}} (Eds.). \bibinfo{publisher}{Springer International
  Publishing}, \bibinfo{address}{Cham}, \bibinfo{pages}{230--253}.
\newblock


\bibitem[\protect\citeauthoryear{Severi, Leek, and Dolan{-}Gavitt}{Severi
  et~al\mbox{.}}{2018}]%
        {DBLP:conf/dimva/SeveriLD18}
\bibfield{author}{\bibinfo{person}{Giorgio Severi}, \bibinfo{person}{Tim Leek},
  {and} \bibinfo{person}{Brendan Dolan{-}Gavitt}.}
  \bibinfo{year}{2018}\natexlab{}.
\newblock \showarticletitle{Malrec: Compact Full-Trace Malware Recording for
  Retrospective Deep Analysis}. In \bibinfo{booktitle}{\emph{Detection of
  Intrusions and Malware, and Vulnerability Assessment - 15th International
  Conference, {DIMVA} 2018, Saclay, France, June 28-29, 2018, Proceedings}}.
  \bibinfo{pages}{3--23}.
\newblock
\urldef\tempurl%
\url{https://doi.org/10.1007/978-3-319-93411-2\_1}
\showDOI{\tempurl}


\bibitem[\protect\citeauthoryear{Tam, Feizollah, Anuar, Salleh, and
  Cavallaro}{Tam et~al\mbox{.}}{2017}]%
        {DBLP:journals/csur/TamFASC17}
\bibfield{author}{\bibinfo{person}{Kimberly Tam}, \bibinfo{person}{Ali
  Feizollah}, \bibinfo{person}{Nor~Badrul Anuar}, \bibinfo{person}{Rosli
  Salleh}, {and} \bibinfo{person}{Lorenzo Cavallaro}.}
  \bibinfo{year}{2017}\natexlab{}.
\newblock \showarticletitle{The Evolution of Android Malware and Android
  Analysis Techniques}.
\newblock \bibinfo{journal}{\emph{{ACM} Comput. Surv.}} \bibinfo{volume}{49},
  \bibinfo{number}{4} (\bibinfo{year}{2017}), \bibinfo{pages}{76:1--76:41}.
\newblock
\urldef\tempurl%
\url{https://doi.org/10.1145/3017427}
\showDOI{\tempurl}


\bibitem[\protect\citeauthoryear{Ugarte-pedrero, Balzarotti, Santos, and
  Bringas}{Ugarte-pedrero et~al\mbox{.}}{[n.d.]}]%
        {Ugarte-pedrero_sok:deep}
\bibfield{author}{\bibinfo{person}{Xabier Ugarte-pedrero},
  \bibinfo{person}{Davide Balzarotti}, \bibinfo{person}{Igor Santos}, {and}
  \bibinfo{person}{Pablo~G. Bringas}.} \bibinfo{year}{[n.d.]}\natexlab{}.
\newblock \bibinfo{title}{SoK: Deep Packer Inspection: A Longitudinal Study of
  the Complexity of Run-Time Packers}.
\newblock
\newblock


\bibitem[\protect\citeauthoryear{Wei, Li, Roy, Ou, and Zhou}{Wei
  et~al\mbox{.}}{2017}]%
        {10.1007/978-3-319-60876-1_12}
\bibfield{author}{\bibinfo{person}{Fengguo Wei}, \bibinfo{person}{Yuping Li},
  \bibinfo{person}{Sankardas Roy}, \bibinfo{person}{Xinming Ou}, {and}
  \bibinfo{person}{Wu Zhou}.} \bibinfo{year}{2017}\natexlab{}.
\newblock \showarticletitle{Deep Ground Truth Analysis of Current Android
  Malware}. In \bibinfo{booktitle}{\emph{Detection of Intrusions and Malware,
  and Vulnerability Assessment}}, \bibfield{editor}{\bibinfo{person}{Michalis
  Polychronakis} {and} \bibinfo{person}{Michael Meier}} (Eds.).
  \bibinfo{publisher}{Springer International Publishing},
  \bibinfo{address}{Cham}, \bibinfo{pages}{252--276}.
\newblock


\bibitem[\protect\citeauthoryear{Yin, Song, Egele, Kruegel, and Kirda}{Yin
  et~al\mbox{.}}{2007}]%
        {Yin:2007:PCS:1315245.1315261}
\bibfield{author}{\bibinfo{person}{Heng Yin}, \bibinfo{person}{Dawn Song},
  \bibinfo{person}{Manuel Egele}, \bibinfo{person}{Christopher Kruegel}, {and}
  \bibinfo{person}{Engin Kirda}.} \bibinfo{year}{2007}\natexlab{}.
\newblock \showarticletitle{Panorama: Capturing System-wide Information Flow
  for Malware Detection and Analysis}. In \bibinfo{booktitle}{\emph{Proceedings
  of the 14th ACM Conference on Computer and Communications Security}}
  \emph{(\bibinfo{series}{CCS '07})}. \bibinfo{publisher}{ACM},
  \bibinfo{address}{New York, NY, USA}, \bibinfo{pages}{116--127}.
\newblock
\showISBNx{978-1-59593-703-2}
\urldef\tempurl%
\url{https://doi.org/10.1145/1315245.1315261}
\showDOI{\tempurl}


\bibitem[\protect\citeauthoryear{Zhou and Jiang}{Zhou and Jiang}{2012}]%
        {Zhou:2012:DAM:2310656.2310710}
\bibfield{author}{\bibinfo{person}{Yajin Zhou} {and} \bibinfo{person}{Xuxian
  Jiang}.} \bibinfo{year}{2012}\natexlab{}.
\newblock \showarticletitle{Dissecting Android Malware: Characterization and
  Evolution}. In \bibinfo{booktitle}{\emph{Proceedings of the 2012 IEEE
  Symposium on Security and Privacy}} \emph{(\bibinfo{series}{SP '12})}.
  \bibinfo{publisher}{IEEE Computer Society}, \bibinfo{address}{Washington, DC,
  USA}, \bibinfo{pages}{95--109}.
\newblock
\showISBNx{978-0-7695-4681-0}
\urldef\tempurl%
\url{https://doi.org/10.1109/SP.2012.16}
\showDOI{\tempurl}


\end{thebibliography}
